\documentclass[a4paper,fleqn]{cas-sc}

\usepackage[numbers]{natbib}
\usepackage{amsmath,amsxtra}
\def\tsc#1{\csdef{#1}{\textsc{\lowercase{#1}}\xspace}}
\tsc{WGM}
\tsc{QE}
\begin{document}
\let\WriteBookmarks\relax
\def\floatpagepagefraction{1}
\def\textpagefraction{.001}

\shorttitle{<High Fusion Computers: The IoTs, Edges, Data Centers, and Humans-in-the-loop as a Computer>}    
%: Building Open-source High Fusion Computers  for Emerging and Future Applications
\shortauthors{<short author list for running head>}  

\title [mode = title]{High Fusion Computers: The IoTs, Edges, Data Centers, and Humans-in-the-loop as a Computer}  
%: Building Open-source High Fusion Computers  for Emerging and Future Applications 

%\tnotemark[<tnote number>] 

%\tnotetext[<tnote number>]{<tnote text>} 

\author[1,2]{Wanling Gao}[%type=editor,
                        %auid=000,bioid=1,
                        %prefix=Sir,
                        %role=Researcher,
                        orcid=0000-0002-3911-9389
                        ]
\ead{gaowanling@ict.ac.cn}
\author[1,2]{Lei Wang}[%type=editor,
                        %auid=000,bioid=1,
                        %prefix=Sir,
                        %role=Researcher,
                        %orcid=0000-0002-3911-9389
                        ]
%\cormark[1]
\ead{wanglei_2011@ict.ac.cn}
%\ead[url]{website information}
\author[1,2]{Mingyu Chen}[%type=editor,
                        %auid=000,bioid=1,
                        %prefix=Sir,
                        %role=Researcher,
                        %orcid=0000-0002-3911-9389
                        ]
%\cormark[1]
\ead{cmy@ict.ac.cn}
\author[1,2]{Jin Xiong}[%type=editor,
                        %auid=000,bioid=1,
                        %prefix=Sir,
                        %role=Researcher,
                        %orcid=0000-0002-3911-9389
                        ]
%\cormark[1]
\ead{xiongjin@ict.ac.cn}
\author[1,2]{Chunjie Luo}[%type=editor,
                        %auid=000,bioid=1,
                        %prefix=Sir,
                        %role=Researcher,
                        %orcid=0000-0002-3911-9389
                        ]
%\cormark[1]
\ead{luochunjie@ict.ac.cn}
\author[1,2]{Wenli Zhang}[%type=editor,
                        %auid=000,bioid=1,
                        %prefix=Sir,
                        %role=Researcher,
                        %orcid=0000-0002-3911-9389
                        ]
%\cormark[1]
\ead{zhangwl@ict.ac.cn}
\author[3]{Yunyou Huang}[%type=editor,
                        %auid=000,bioid=1,
                        %prefix=Sir,
                        %role=Researcher,
                        %orcid=0000-0002-3911-9389
                        ]
%\cormark[1]
\ead{huangyunyou@gxnu.edu.cn}
\author[4]{Weiping Li}[%type=editor,
                        %auid=000,bioid=1,
                        %prefix=Sir,
                        %role=Researcher,
                        %orcid=0000-0002-3911-9389
                        ]
%\cormark[1]
\ead{weiping2012c@gmail.com}
\author[1,2]{Guoxin Kang}[%type=editor,
                        %auid=000,bioid=1,
                        %prefix=Sir,
                        %role=Researcher,
                        %orcid=0000-0002-3911-9389
                        ]
%\cormark[1]
\ead{kangguoxin@ict.ac.cn}
\author[5]{Chen Zheng}[%type=editor,
                        %auid=000,bioid=1,
                        %prefix=Sir,
                        %role=Researcher,
                        %orcid=0000-0002-3911-9389
                        ]
%\cormark[1]
\ead{zhengchenjason@hotmail.com}
\author[1,2]{Biwei Xie}[%type=editor,
                        %auid=000,bioid=1,
                        %prefix=Sir,
                        %role=Researcher,
                        %orcid=0000-0002-3911-9389
                        ]
%\cormark[1]
\ead{xiebiwei@ict.ac.cn}
\author[1,2]{Shaopeng Dai}[%type=editor,
                        %auid=000,bioid=1,
                        %prefix=Sir,
                        %role=Researcher,
                        %orcid=0000-0002-3911-9389
                        ]
%\cormark[1]
\ead{daishaopeng@ict.ac.cn}
\author[6]{Qian He}[%type=editor,
                        %auid=000,bioid=1,
                        %prefix=Sir,
                        %role=Researcher,
                        %orcid=0000-0002-3911-9389
                        ]
%\cormark[1]
\ead{heqian@bosc.ac.cn}
\author[6]{Hainan Ye}[%type=editor,
                        %auid=000,bioid=1,
                        %prefix=Sir,
                        %role=Researcher,
                        %orcid=0000-0002-3911-9389
                        ]
%\cormark[1]
\ead{yehainan@bosc.ac.cn}
\author[1,2]{Yungang Bao}[%type=editor,
                        %auid=000,bioid=1,
                        %prefix=Sir,
                        %role=Researcher,
                        %orcid=0000-0002-3911-9389
                        ]
%\cormark[1]
\ead{baoyg@ict.ac.cn}
\author[1,2]{Jianfeng Zhan}[%type=editor,
                        %auid=000,bioid=1,
                        %prefix=Sir,
                        %role=Researcher,
                        %orcid=0000-0002-3911-9389
                        ]
\cormark[1]%{Jianfeng Zhan is the corresponding author.}
\ead{zhanjianfeng@ict.ac.cn}

%\credit{Conceptualization of this study, 
%          Methodology, Software}

\affiliation[1]{organization={State Key Lab of Processors, and Research Center for Advanced Computer Systems, Institute of Computing Technology, Chinese Academy of Sciences},
                addressline={No. 6 Kexueyuan South Road, Haidian District}, 
                %city={Beijing},
%               citysep={}, % Uncomment if no comma needed 
%                               between city and postcode
                postcode={100190}, 
                state={Beijing},
                country={China}}
\affiliation[2]{organization={University of Chinese Academy of Sciences},
                addressline={No. 19 (A) Yuquan Road, Shijingshan District}, 
                %city={Beijing},
%               citysep={}, % Uncomment if no comma needed 
%                               between city and postcode
                postcode={100049}, 
                state={Beijing},
                country={China}}
\affiliation[3]{organization={Guangxi Normal University},
                addressline={No. 15, Yucai Road, Qixing District}, 
                city={Guilin},
%               citysep={}, % Uncomment if no comma needed 
%                               between city and postcode
                postcode={541010}, 
                state={Guangxi},
                country={China}}
\affiliation[4]{organization={Civil Aviation Flight University of China},
                addressline={No. 46, Section 4, Nanchang Road}, 
                city={Guanghan},
%               citysep={}, % Uncomment if no comma needed 
%                               between city and postcode
                postcode={618307}, 
                state={Sichuan},
                country={China}}
\affiliation[5]{organization={Institute of Software, Chinese Academy of Sciences},
                addressline={No. 4, South 4th Street, Zhongguancun}, 
                %city={Beijing},
%               citysep={}, % Uncomment if no comma needed 
%                               between city and postcode
                postcode={100190}, 
                state={Beijing},
                country={China}}
\affiliation[6]{organization={Beijing Institute of Open Source Chip},
                addressline={No. 7 Caoqiao, northwest corner of Haidian Bridge, Haidian District}, 
                %city={Beijing},
%               citysep={}, % Uncomment if no comma needed 
%                               between city and postcode
                postcode={100080}, 
                state={Beijing},
                country={China}}

%\author[<aff no>]{<author name>}[<options>]
%\author[<aff no>]{<author name>}

%\cormark[1]{Jianfeng Zhan is the corresponding author.}
\cortext[cor1]{Corresponding author is Jianfeng Zhan.}

%\fnmark[<footnote mark no>]

%\ead{<email address>}

%\ead[url]{<URL>}

%\credit{<Credit authorship details>}

%\affiliation[<aff no>]{organization={},
%          citysep={}, % Uncomment if no comma needed between city and postcode

%\author[<aff no>]{<author name>}[<options>]
%\author[<aff no>]{<author name>}

% Footnote of the second author

% Email id of the second author
%\ead{}



%\affiliation[<aff no>]{organization={},
%%          citysep={}, % Uncomment if no comma needed between city and postcode

%\cortext[1]{Corresponding author}



\begin{abstract}
%\ zhan jianfeng
Emerging and future applications rely heavily upon systems consisting of Internet of Things (IoT), edges, data centers, and humans-in-the-loop. Significantly different from warehouse-scale computers that serve independent concurrent user requests, this new class of computer systems directly interacts with the physical world, considering humans an essential part and performing safety-critical and mission-critical operations;   their computations have intertwined dependencies between not only adjacent execution loops but also actions or decisions triggered by IoTs, edge, datacenters, or humans-in-the-loop; the systems must first satisfy the accuracy metric in predicting, interpreting, or taking action before meeting the performance goal under different cases.
%This new class of systems demands resources that are several orders of magnitude beyond the reach of the state-of-the-practice systems. 

This article argues we need a paradigm shift to reconstruct the IoTs, edges, data centers, and humans-in-the-loop as a computer rather than a distributed system. We coin a new term, high fusion computers (HFCs), to describe this class of systems. The fusion in the term has two implications:  fusing IoTs, edges, data centers, and humans-in-the-loop as a computer, fusing the physical and digital worlds through HFC systems. HFC is a pivotal case of the open-source computer systems initiative. We laid out the challenges, plan, and call for uniting our community’s wisdom and actions to address the HFC challenges. Everything, including the source code, will be publicly available from the project homepage: ~\url{https://www.computercouncil.org/HFC/}.
%Second, under different conditions like the best-case, worst-case, or average-case, performance, quality~\footnote{The quality metric measures the accuracy of an application, task, or algorithm in predicting and interpreting.}, and interpretability~\footnote{ According to ~\cite{miller2019explanation}, interpretability indicates the degree to which an observer can understand the cause of a decision.} of computation results are equally critical.  

%\ zhan jianfeng
\end{abstract}


%\item 
%\item 
%\item 

% Keywords
% \begin{keywords}
%  \sep Emerging and future applications \sep Safety-mission \sep Mission-critical \sep IoT \sep Edge \sep Datacenter \sep Humans-in-the-loop  \sep  Open-source computer systems \sep High Fusion Computer \sep HFC 
 
% % The IoTs, Edges, Datacenters, Networks and Humans-in-the-loop as a Computer 
% \end{keywords} %commented by zhengxin
 % update: Safety-mission -> Safety-critical

\maketitle

% Main text
%\section{Introduction}\label{sec-intro}

% Use the style of numbering in square brackets.
% If nothing is used, default style will be taken.
%\begin{enumerate}[a)]
%\item 
%\item 
%\item 
%\end{enumerate}  

%\begin{itemize}
%\item 
%\item 
%\item 
%\end{itemize}  

%\begin{description}
%\item[]
%\item[] 
%\item[] 
%\end{description}  

% Figure
%\begin{figure}[<options>]
%	\centering
%		\includegraphics[<options>]{}
%	  \caption{}\label{fig1}
%\end{figure}

%\begin{table}[<options>]
%\caption{}\label{tbl1}
%\begin{tabular*}{\tblwidth}{@{}LL@{}}
%\toprule
%  &  \\ % Table header row
%\midrule
% & \\
% & \\
% & \\
% & \\
%\bottomrule
%\end{tabular*}
%\end{table}

% Uncomment and use as the case may be
%\begin{theorem} 
%\end{theorem}

% Uncomment and use as the case may be
%\begin{lemma} 
%\end{lemma}


\section{Introduction}

%\ zhan jianfeng
The past decades have witnessed solid achievements and ambitious plans on planetary-scale infrastructures.  Typical examples include but are not only limited to Grid computing~\cite{foster2008cloud},  planet-scale data centers hosting internet services or called warehouse-scale computers~\cite{GCP2016_google},  virtual supercomputers in the cloud~\cite{Mike2022VSC}, interplanetary Internet~\cite{burleigh2003interplanetary},  networked systems of embedded computers~\cite{national2001embedded}, planet-scale computing networks~\cite{xu2022information,wang2021building} or planetary computer~\cite{MSR_pc}. However, emerging and future applications raise daunting challenges beyond the reach of the state-of-the-practice systems.

Emerging and future applications rely heavily on systems consisting of IoTs, edges, data centers, and humans-in-the-loop~\cite{nunes2015survey}. 
%, which raises daunting challenges beyond the reach of the state-of-the-practice systems. 
These networked systems of embedded computers~\cite{national2001embedded} or IoTs, collaborating with data centers, edges, and humans-in-the-loop, can radically change the way people interact with the physical world and perform safety-critical~\footnote{According to ~\cite{sommerville2011software}, a safety-critical system is a  system whose failure may result in injury, loss of life or serious environmental damage, e.g., a control system for "a chemical manufacturing plant". ~\url{https://ifs.host.cs.st-andrews.ac.uk/Books/SE9/Web/Dependability/CritSys.html}} or mission-critical~\footnote{According to ~\cite{sommerville2011software}, a mission-critical system is "a system whose failure may fail some goal-directed activity, e.g., a navigational system for a spacecraft"; a business-critical system is "a system whose failure may result in very high costs for the business using that system, e.g., customer accounting system in a bank".   ~\url{https://ifs.host.cs.st-andrews.ac.uk/Books/SE9/Web/Dependability/CritSys.html}} tasks. This new class of  systems appears in many forms and continues to expand:  "implemented as a kind of digital nervous system to enable instrumentation of all sorts of spaces, ranging from in situ environmental monitoring to surveillance of battlespace conditions"~\cite{national2001embedded}, e.g., climate change monitoring and defense systems; embodied as integrated  instrumentation, operation, maintenance, and regulation facilities of critical physical infrastructure, e.g., industrial digital twin, 
energy infrastructure management and civil aviation regulation; "employed in personal monitoring strategies (both defense-related and civilian), synthesizing information from sensors on and within a person with information from laboratory tests and other  sources"~\cite{national2001embedded}, e.g., medical emergency applications;   augmented as an extension of 
the real-world life for entertainment, education, and social activities, e.g., Metaverse; instrumented as a kind of digital sensing and autonomic control systems to perform safety-critical or mission-critical tasks, e.g., automotive driving and interplanetary explorations. 

%dramatically
%affect scientific data collection capabilities, ranging from new techniques
%for precision agriculture and biotechnological research to detailed environmental and pollution monitoring.
%or even augmenting the life.

Significantly different from warehouse-scale computers that non-stop serve independent concurrent user requests~\cite{GCP2016_google}, the new class of computer systems has three unique requirements.  
First, they directly interact with the physical world -- considering humans an essential part: human-in-the-loop, performing safety-critical and mission-critical operations, and a significant fraction of actions may have an irreversible effect. The role of humans and their interactions with the other system components can not be ignored in the final impact on the physical world. 
%Second, under different conditions like the best-case, worst-case, or average-case, performance, quality~\footnote{The quality metric measures the accuracy of an application, task, or algorithm in predicting and interpreting.}, and interpretability~\footnote{ According to ~\cite{miller2019explanation}, interpretability indicates the degree to which an observer can understand the cause of a decision.} of computation results are equally critical.  
Second, unlike Internet services that process independent concurrent requests across data centers,  their computations have intertwined dependencies between not only adjacent execution loops (which we call internal dependencies) but also actions or decisions triggered by IoTs, edge, data centers, or humans-in-the-loop (which we call external dependencies), and traverse different paths through and around IoTs, edges, and data centers. Third, under this highly entangled state, the systems must first satisfy the quality metric before meeting the performance goal under different conditions, like worst-case, average-case, and best-case. The quality metric measures the accuracy of an application, task, or algorithm in predicting, interpreting, or taking action.

Even considering a simple   IoT
application -- 95\% queries reporting the status and 5\% user queries accessing the database,
serving 10-billion devices needs about one thousand to one hundred thousand nodes under the threshold of 50 milliseconds response time. Further, considering the three unique system requirements mentioned above and the sea change in computing, data access, and networking patterns, this new class of computer systems demand resources that are several orders of magnitude beyond the reach of the state-of-the-practice systems, which raises daunting challenges. 
%Considering the two unique requirements mentioned above and other factors
%Let's consider sophisticated functions like object recognition interference, OODA loop~\cite{richards2020boyd} or even complex Avatar behavior. The gap will be several orders of magnitude beyond the reach of the state-of-the-practice systems. 

%, which are three or five orders of magnitude beyond the reach of the state-of-the-practice systems.
%read and write accesses to the Redis database, 

 %, and represents a 417% performance increase in just two years. When compared to the other CPU-based runs, this one sits at position 20. The GPU-based runs are certainly impressive, but ranking them separately makes for the best apples-to-apples comparison.

\begin{figure*}
	\centering
		\includegraphics[scale=0.65]{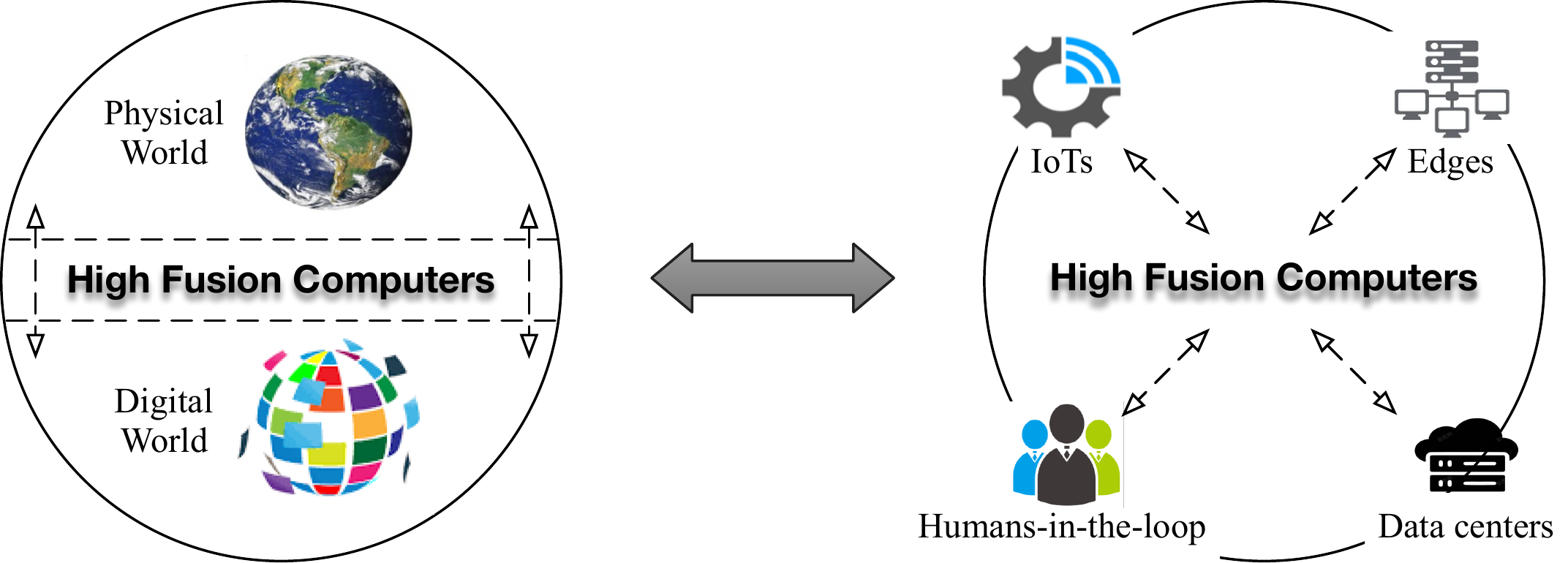}
	\caption{A Concept Viewpoint of High Fusion Computers (HFCs).}
	\label{hfc-twofusion}
\end{figure*}

This article argues that we need a paradigm shift to rebuild the IoTs, edges, data centers, and humans-in-the-loop as a computer rather than a distributed system. We coin a new term,  high fusion computers (HFCs), to describe this new class of computer systems. According to the Oxford English Dictionary, fusion means "the process or result of joining two or more things together to form a single entity." Fusion in HFCs has two-fold meanings: fusing IoTs, edges, data centers, and humans-in-the-loop as a computer, fusing the physical world and digital world. Fig.~\ref{hfc-twofusion} shows a concept viewpoint of HFCs. Our intuition in rebuilding the HFC systems is simple: We explicitly value the role of humans-in-the-loop in different contexts and consider them as essential components of the system; we aggressively embrace co-design from vertical and horizontal dimensions, and will co-explore the design space from the algorithms,  runtime systems, resource management, storage, memory, networking, and chip systems from a vertical dimension; Meanwhile, we will consider the close collaboration among IoTs, Edges, data centers, and humans-in-the-loop from a horizontal dimension, and ponder how to facilitate the interactions of humans-in-the-loop with other hardware and software systems. 

%This article argues that we need a paradigm shift to reconstruct the IoTs, edges, datacenter, and networks as a computer rather than a distributed system. We coin a new term, a planet-scale computer (PSC), to describe this system. Table~\ref{comparition_table}  summarizes the unique requirements of PSC from other Planet-scale infrastructures. Our intuition is simple; we will aggressively embrace co-design from vertical and horizontal dimensions. We will co-explore the design space from the algorithms,  runtime systems, resource management, storage, memory, networking, and chip systems from a vertical dimension. We will consider the close collaboration among IoTs, Edges, and datacenter from a horizontal dimension.

 To dismantle the complexity of building the systems and improve the efficiency, we use the funclet abstraction, architecture, and methodology, inspired by the philosophy of building large systems out of smaller functions~\cite{moore1965cramming, ZHAN2022OSSystem}. The funclet abstract represents the common proprieties of basic building blocks:  each funclet is "a well-defined, independently deployable,  testable, evolvable, and reusable functionality with modest complexity; funclets interoperate with each other through well-defined interconnections"~\cite{ZHAN2022OSSystem}. Four kinds of funclets form the four-layer funclet architecture: chiplet, HWlet, envlet, and servlet -- at the chip, hardware, environment management, and service layers, respectively~\cite{ZHAN2022OSSystem}.
 
 %~\cite{ZHAN2022OSSystem}. 
 
We take HFC as a pivotal example of the open-source computer system plan.
 We abstract reusable functions (funclets) across system stacks among IoTs, edges, and data centers. Based on funclets, we rebuild the IoTs, edges, data centers, and humans-in-the-loop as a computer in a structural manner, with full-fledged functions of autonomic resource discovery, management, programming, workload scheduling, and coordinated collaboration between software, hardware and human components.
 Our plans are three-fold.
First, we value the importance of benchmarks and funclet-based standards in evaluating and building the systems.  Second, we emphasize the methodology and tool to facilitate the workload-driven exploration of the system and architecture design space. 
 Third, we will provide the first open-source implementation of the funclet architecture of HFC systems.  
 
 %Third, we provide open-source EDA, FPGA, and other emulation tools for designing and verifying the funclet architecture. 

 %servlet, Envlet, HWlet, and chiplet across the IoT, edge, and datacenter architectures. 
 %aim to build PSCs that provide unified management and efficient collaboration of the IoTs, edges, networks, and datacenters, and automatically schedule the optimal execution strategy for emerging and future applications.

  We organize the rest of this paper as follows. Section 2 explains the motivation. Section 3 illustrates the HFC challenges. Section 4 explores the HFC software and hardware design space. Section 5 describes our plan. Section 6 summarizes the related work. Section 7 concludes.

 %Fourth, we provide the open-source implementation of  HWlet, Envlet, and Servlet architecture. 

%\ zhan jianfeng

%What is a PSC system?

%Why we build new PSC systems

%The challenges

%The plans

%Another goal is to tackle information technology decoupling.

% Our solutions.  architecture, standards, innovations

\section{Motivation}~\label{motivation}

In this section, we first analyze seven typical emerging and future applications' unique requirements, then explain why we need to build an HFC system.

%\textbf{list some methodology.}

\subsection{The requirements of emerging and future applications} ~\label{motivation_requirement}

\begin{table*}[htbp]
\renewcommand\arraystretch{1.2}
\scriptsize
\centering
\caption{Summarization of emerging or future applications.}\label{comparition_table}
\center
\begin{tabular}{|p{0.52in}|p{0.51in}|p{0.28in}|p{0.41in}|p{0.58in}|p{0.8in}|p{0.7in}|p{0.7in}|p{0.6in}|}
%\begin{tabular}{|p{0.6in}|p{0.65in}|p{0.5in}|p{0.55in}|p{1in}|p{0.8in}|p{1.2in}|}
%\begin{tabular}{|p{0.9in}|p{0.63in}|p{0.58in}|p{0.55in}|p{0.55in}|p{0.55in}|p{0.4in}|p{0.42in}|}
\hline

Application & Critical or typical tasks & Task type & Effect & Metrics & IoT Devices & Dependencies$^*$ & Data Management & Access Patterns \\
 \hline
 \multirow{6}{0.6in}{Medical emergency management} & Emergency detection & Safety-critical & Reversible & Worst-case & \multirow{6}{0.9in}{camera, blood pressure monitor, spirometer, gyroscopes, CT scanner, mass spectrometer, etc.} & \multirow{6}{0.8in}{Observe, fuse, recommend, train (Internal \& external dependencies)} & \multirow{6}{0.7in}{image, video, relational data, XML}  & \multirow{6}{0.7in}{Real-time write; Non-periodic random read} \\
\cline{2-5}
 & Rescue planning &Mission-critical &Reversible & Average-case& & & &\\
 \cline{2-5}
 & Rapid diagnosis &Safety-critical &Reversible & Worst-case& & & &\\

\hline
\multirow{6}{0.6in}{Autonomous driving} & Trajectory planning & Safety-critical & \multirow{6}{0.41in}{Irreversible} & \multirow{6}{0.58in}{Worst-case} & \multirow{6}{0.9in}{Camera, LiDAR, Radar, ultrasonic, GNSS, GPS, etc.} & \multirow{6}{0.8in}{Observe, fuse, act, coordinate (Internal \& external dependencies)} & \multirow{6}{0.8in}{image, video, LAS binary, ASCII~\cite{chen2007airborne}, text, XML, float matrix, csv }
& \multirow{6}{0.7in}{Real-time write; Periodic burst read} \\
\cline{2-3}
 & Surrounding object detection &Safety-critical & & & & & &\\
 \cline{2-3}
 & Autonomic control & Mission-critical & & & & & &\\
 
\hline
Smart defense systems &  battlespace surveillance& Safety-critical & Irreversible & Worst-case  & Seismic, acoustic, magnetic, and imaging sensors or terminal control units, etc. & Observe, orient, decide, act (Internal \& external dependencies)  & SEG-Y files, MP3, WAV, image & Real-time write; Real-time read  \\

\hline
\multirow{6}{0.6in}{Digital Twin} & Smart manufacturing& Mission-critical & Reversible & Average-case & \multirow{6}{0.9in}{Cameras, sensors, analog-to-digital converter, digital-to-analog converter, etc.}  & \multirow{6}{0.8in}{Observe, model, decide, control (Internal \& external dependencies)} & \multirow{6}{0.7in}{image, binary, relational data} & \multirow{6}{0.7in}{Periodic write; Periodic read} \\
\cline{2-5}
 &Oil well drilling &Safety-critical & Irreversible &Worst-case & & & &\\
 & & & & & & & &\\
  & & & & & & & &\\

\hline
\multirow{6}{0.6in}{Civil aviation safety regulation} & Airport security & Safety-critical & \multirow{6}{0.41in}{Irreversible} & \multirow{6}{0.58in}{Worst-case}  & \multirow{6}{0.9in}{Cameras, Radars, VHF, Flight-Data Acquisition Unit (FDAU), etc.} & \multirow{6}{0.8in}{Observe, fuse, decide, and alert (Internal \& external dependencies)} & \multirow{6}{0.7in}{image, binary, relational data}
& \multirow{6}{0.7in}{Real-time write; Random read} \\
\cline{2-3}
 & Air navigation & Mission-critical & & & & & & \\
 \cline{2-3}
 & Anomaly detection & Safety-critical & & & & & & \\

\hline
\multirow{6}{0.6in}{Metaverse} & Scenario generating & Mission-critical & Reversible & Avearge-case  & \multirow{6}{0.9in}{Head-mounted display (HMD), Handheld devices (HHDs), etc.} & \multirow{6}{0.8in}{Observe, recognize, fuse, act (Internal \& external dependencies)} & \multirow{6}{0.7in}{dynamic multimedia, relational data} & \multirow{6}{0.7in}{Real-time write; Real-time read} \\
\cline{2-5}
 & Avatar maintaining &Mission-critical & Reversible & Avearge-case & & & & \\
 \cline{2-5}
 & Decentralised finance & Mission-critical & Irreversible & Worst-case&  & & & \\

\hline
\multirow{8}{0.6in}{Interplanetary explorations} & Knowledge discovery & Mission-critical  &  Reversible & Best-case & \multirow{8}{0.9in}{Satellite, Space probe, Robots, etc.} & \multirow{8}{0.8in}{Observe, recognize, infer, control (Internal \& external dependencies)} &  \multirow{8}{0.8in}{image, Hierarchical data format(HDF), Network Common Data Form (NetCDF)}& \multirow{8}{0.7in}{Real-time write; Batch transfer; Random read} \\
\cline{2-5}
 & Collision avoidance & Safety-critical & Irreversible & Worst-case & & & & \\
 & & & & & & & &\\
 \cline{2-5}
 & Space navigation & Mission-critical & Irreversible & Worst-case & & & &\\
 & & & & & & & &\\

\hline

\multicolumn{9}{l}{$^*$Internal dependency indicates the dependencies between adjacent execution loops. } \\
\multicolumn{9}{l}{$^*$External dependency indicates the dependency between actions or decisions triggered by IoTs, edge, data centers, or humans-in-the-loop.}\\

\end{tabular}
\end{table*}

This subsection analyzes seven emerging and future applications. Table 1 characterizes those applications. I detail two applications specifically as follows.

\subsubsection{Medical emergency management}  %//yunyou huang

According to the data from the World Bank, there are more than 723 million people over the age of 65 in the world in 2020, accounting for 9.321\% of the world's total population~\cite{2020population}. 
What's worse, despite the slowdown in world population growth, the proportion of people over the age of 65 is growing rapidly, which will account for 16\% of the total population by 2050~\cite{desa2019world,teixeira2021wearable}. Due to the decline of physical function and pathological changes, the elderly will experience many unplanned emergencies in terms of companionship, nursing, medical treatment, etc., bringing massive pressure to the emergency medical care of the entire lifecycle in the future~\cite{pal2018internet}. Furthermore, when
disasters occur, the surge of patients can quickly overwhelm the already overcrowded care facilities~\cite{ko2010medisn}.

%The huge elderly population puts enormous pressure on the healthcare system, especially in the emergency medical departments. 

\begin{figure}[h]
\centering
\includegraphics[width=9cm,height=9cm]{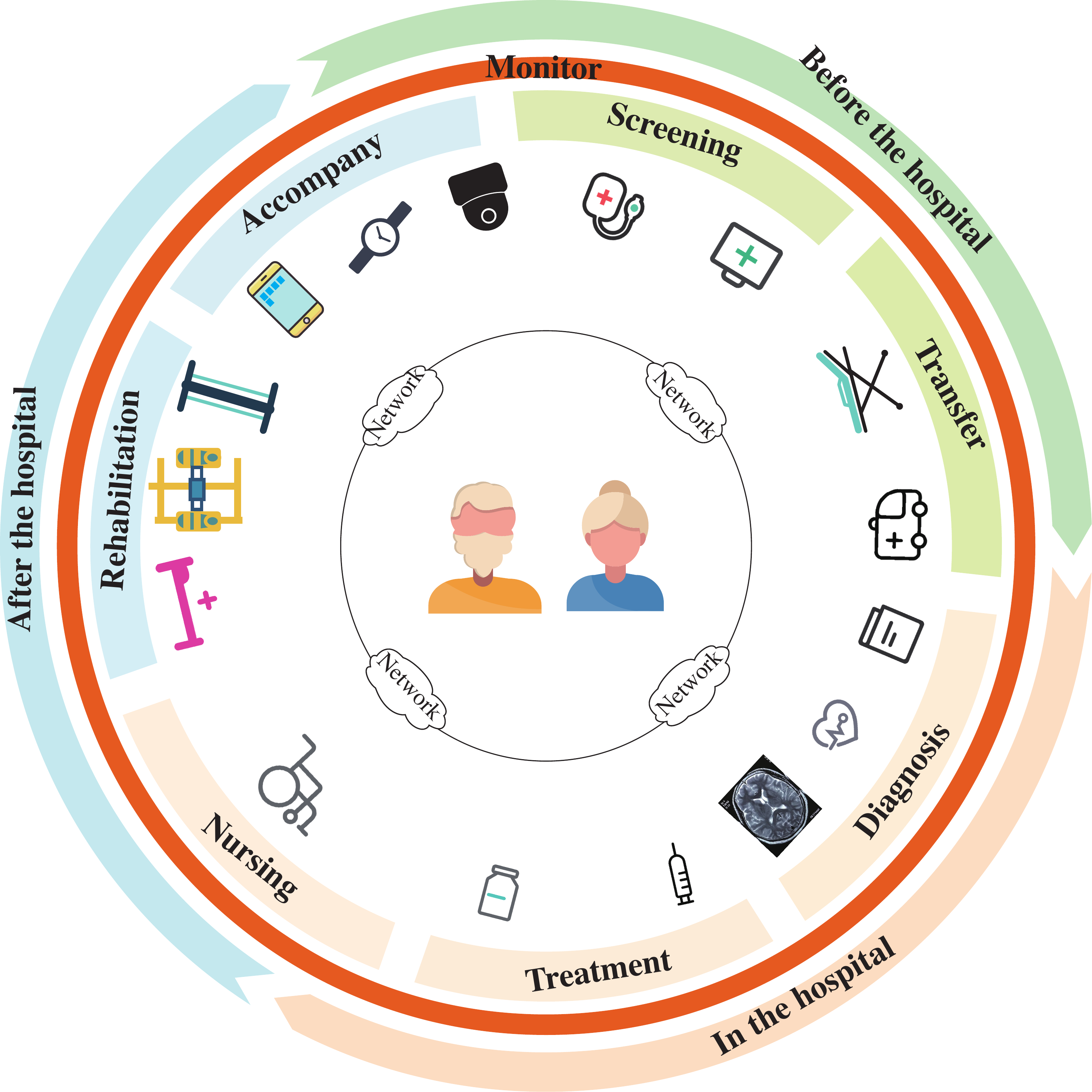}
\caption{The healthcare lifecycle of the elderly.}
\label{lifecycle-healthcare}
\end{figure}

%As a typical safety-critical system, the medical emergency management system is  ``employed in personal monitoring strategies (both defense-related and civilian), integrating information from sensors on and within a person with information from laboratory tests and other sources”~\cite{national2001embedded}.
Many computing technologies (e.g., IoT, AI, cloud computing) are introduced to support the health system to overcome current and future dilemmas of the elderly emergency medical care\cite{balasubramaniam2019iot,alexandru2019iot,majumder2017smart,liu2019novel}. However, as shown in  
Figure~\ref{lifecycle-healthcare}, Many elderly emergency medical care issues of the entire lifecycle remain unsolved, posing enormous challenges to the computer systems sustaining emergency medical care applications.

\begin{itemize}
\item \textbf{Task types: emergent and mainly safe-critical.} The elderly are more likely to experience medical emergencies (e.g., stroke, myocardial infarction, falls, etc.) than younger adults, and these events often cause more significant harm to the elderly. When an emergency medical event occurs, it is required that the medical system can handle the safe-critical task in real-time and that the medical system can reasonably allocate medical resources for rapid rescue. It is worth noticing that medical experts play a decisive role in the system. Medical emergency management systems consider medical professionals a reliable external component in the control loop, which we call reliable-human-in-the-loop. In this scenario, the system may make recommendations, but the medical expert takes the responsibility, and the decision made by the system is Reversible.

  \item \textbf{Metrics:} In the worst-case and average-case, the quality of computation results is vital. Though the medical experts will take the final responsibility,  the systems are valuable only by providing high-quality and interpretable computation results. To provide real-time and safe-critical services, the worst-case performance is important besides the average performance, including the latency and throughput. Since the workloads are often spiky, the systems must gracefully handle overloading.   
  
   Due to the specificity of medical care, security and privacy are always the first issues that medical systems have to consider.  The future healthcare systems involve a more significant number and variety of devices, a larger population, and more applications, and its complexity brings more significant challenges to security and privacy.

  %Due to the specificity of medical care, security and privacy are always the first issues medical systems have to consider.  The future healthcare systems involve a more significant number and variety of devices, a larger population, and more applications, and its complexity brings more significant challenges to security and privacy.
  
  %Under different conditions like best-case, worst-case, or average-case, the quality and interpretability of computation results are equally critical in addition to the performance.

\item \textbf{Various IoT devices generate a massive volume of heterogeneous data.} Unlike traditional emergency medical care, modern elderly emergency medical care has been extended to their daily lives: before, during, and after the hospital.   The emergency medical systems not only rely on a large number of different sensors (such as cameras, gyroscopes, blood pressure monitors, etc.) and also need to access various professional medical equipment (such as CT scanner, mass spectrometer, spirometer, etc.). These devices generate a large amount of heterogeneous data. The emergency medical care system needs to integrate and process large amounts of heterogeneous data in real-time to provide emergency medical services to the elderly throughout their life cycle.

\item \textbf{Computation patterns, computation dependencies, and interaction patterns:}

%Based on the physiological characteristics of the elderly, the healthcare system needs to provide medical services throughout the life cycle, and a large number of different applications (such as diagnosis, monitoring, treatment, etc.) and a large number of different systems (such as smart home, smart city, medical imaging system, etc.) need to work together.

The computation patterns follow the observe, fuse, recommend, and train patterns.  
The IoT devices observe the data of the patients at different levels. The system may fuse various observations at the edge or data center.  The data centers or edges will train and update an AI model through the widely collected and labeled data. The IoT or edge makes a recommendation like alert and further-taken actions. The medical experts make a final decision.

The computation dependencies are mainly internal dependencies -- the dependency between adjacent execution loops of observing, fusing, recommending, and training. For example, a patient's previous diagnosis and treatment recommendations would impact their subsequent recommendations. Besides, external dependencies exist between the decisions triggered by different IoTs, edges, data centers, or humans. Typical examples include new emergency cases reported by the IoTs, the newly trained models, and interventions brought out by the experts. 

Over the other applications, the interaction patterns are simpler. Each IoT works within different conditions, and each computation may trigger different algorithms but only involve local data. Recommendations may be made at edges locally, involving collected data with different spatial-temporal scopes. When training the model, the data are widely collected from IoTs or edges and annotated at the data center for further training. Or in another manner, the labeled data at IoT or edges are distributed training using federation learning techniques~\cite{liang2020flbench,huang2022training}. Then lightweight models are deployed at IoT and edges.

%IED systems have complex computation patterns. Under different scenarios, computation wildly varies from simple status reports, object recognition interference, a much more intricate pattern like OODA (observe, orient, decide, and act)~\cite{richards2020boyd} or even complex Avatar behavior like big data analytics and machine learning.  Last but not least, the IED systems have complex and diverse interaction patterns.  Each IoT works within different conditions. Each computation may trigger different algorithms, involve data with different spatial or temporal scopes, and traverse different paths through and around IoTs, edges, and datacenters. 
 
 \item \textbf{Data management and access pattern:}
 Patients generate a large number of real-time data from various sensors and professional medical devices.
 The formats of patient data are diverse - image, video, relational data, XML, etc~\cite{pollard2018eicu}.
 Patient data is written into the emergency medical care system in real-time.
 When an emergency medical event is detected, the emergency medical care system immediately initiates rapid diagnosis and develops rescue planning.
 Patient data helps medical experts understand how a critical health event unfolds, uncover the geographic characteristics of events, and locate the nearest medical resources.
 Patient data is accessed only when the patient experiences a medical emergency.
 Therefore, medical emergencies result in non-periodic random access patterns.
 %The various sensors and professional medical equipment data are divided into three dimensions: stateful, spatio, and temporal data.
 %When an emergency medical event occurs, the Spatio-temporal data are crucial for the medical experts to allocate medical resources. The temporal data help understand how a critical healthy event evolves. Meanwhile, spatial data uncover the geolocation properties of events and can help locate the closest medical resources in the proximity. The emergency medical events produce non-periodic burst access patterns.
 %The  Stateful data refers to the data that describes the status of service objects.
 %In the medical emergency management system, stateful data describes the physiological index of patients (such as blood pressure, blood glucose, etc.).
 % And the spatio-temporal data contains locations of medical resources and patients change over time and helps patients obtain the medical resources closest to them.

% \item \textbf{Security and privacy:}

\end{itemize}

\subsubsection{Autonomous driving}

%Future vehicles integrate state-of-the-art sensing, communications, artificial intelligence
%s, and automated technologies. These technologies will integrate autonomous vehicles, users, and infrastructure (roads and traffic lights). Millions of sensors and cameras will be deployed on the vehicles and roads. These IoT devices collect vehicles and road information. Some information is processed immediately in the local car and helps make decisions, e.g., decelerate, accelerate or change lanes. Some information is transferred to edge devices or datacenter for further analysis, e.g., traffic Congestion analysis or route planning. These networked systems of IoT on vehicles and roads, collaborating with datacenters through many edge devices, will improve the safety and efficiency of future transportation. 

Autonomous driving is a promising technology that changes the way people travel. According to the standard of SAE International~\cite{shadrin2019analytical}, autonomous driving is classified into six levels---"no driving automation (Level 0), driver assistance (Level 1), partial driving automation (Level 2), conditional driving automation (Level 3), high driving automation (Level 4), and full driving automation (Level 5)"~\cite{shadrin2019analytical}. With its continuous development, self-driving cars will hit the roads and enter into a highly-automated era in the future~\cite{autodriventtdata}.

\begin{itemize}
\item \textbf{Task types: highly-automated and mainly safety-critical.} The future autonomous driving would be highly-automated, even fully-automated, and consider no human in the control loop. The corresponding system needs to perceive and collect multi-source and multi-dimensional data in real-time and respond within several milliseconds. Considering the properties of high autonomy, hard real-time, and potentially destructive effects, the decision and action made by the system are irreversible. 
It will be a system failure in hard real-time when missing a deadline. In auto-driving, missing a deadline will be catastrophic. 
%It is a highly real-time and safety-critical application scenario, pursuing both low latency and high accuracy since a slight mistake would cause heavy casualties.

%The elderly are more likely to experience medical emergencies (e.g., stroke, myocardial infarction, falls, etc.) than younger adults, and these events often cause more significant harm to the elderly. When an emergency medical event occurs, it is required that the medical system can handle the safe-critical task in real-time and that the medical system can reasonably allocate medical resources for rapid rescue. It is worth noticing that medical experts play a decisive role in the system. Medical emergency management systems consider humans a reliable external component in the control loop, which we call reliable-human-in-the-loop.  In this scenario, the system may make recommendations, but the medical expert takes the responsibility, and the decision made by the system is revocable.

\item \textbf{Metrics:} 
In the worst-case and average-case, the quality of computation results is vital as no person takes responsibility. The systems must provide high-quality and interpretable computation results. The worst-case performance of autonomous driving is highly significant. For the safety of cars, pedestrians, and surroundings, an autonomous driving system is demanded to manage and coordinate massive self-driving cars synchronously, assure the performance of almost all vehicles and guarantee the worst-case performance --- make the tail latency as low as possible. 

Security and privacy are critical challenges in autonomous driving. Besides the traditional security issue, security also means the car can perceive the environment, make decisions, and take actions correctly.
%result in serious accidents.

%As the tasks are emergent and safe-critical, in addition to the average performance, the worst-case performance is of paramount importance, including the latency and throughput.   Considering the workloads are often spiky, the systems must gracefully handle overloading.   In the worst-case and average-case, the quality of computation results is also vital. Though the medical experts will take the final responsibility,  the systems are valuable only by providing high-quality and interpretable computation results.

\item \textbf{Various IoT devices generate a massive volume of heterogeneous data.} Autonomous driving depends on a large number of sensors; even a single car may deploy multiple kinds of sensors~\cite{yeong2021sensor,vargas2021overview} including cameras, ultrasonic radar, millimeter-wave radar, lidar,  IMU (Inertial Measurement Unit), etc., to obtain the environment information comprehensively. The input data are multi-source and heterogeneous; thus, the system should be able to fuse multi-sensor data for quick processing.  

%Unlike traditional emergency medical care, modern elderly emergency medical care has been extended to their daily lives: before, during, and after the hospital.   The emergency medical systems not only rely on a large number of different sensors (such as cameras, gyroscopes, blood pressure monitors, etc.) and also need to access various professional medical equipment (such as CT scanner, mass spectrometer, spirometer, etc.). These devices generate a large amount of heterogeneous data. The emergency medical care system needs to integrate and process large amounts of heterogeneous data in real-time to provide emergency medical services to the elderly throughout their life cycle.

\item \textbf{Computation patterns, computation dependencies, and interaction patterns:} 

The computation patterns follow the observe, fuse, act, coordinate, and train patterns.  
The IoT devices observe and detect the data of the weather, surroundings, road lanes, traffic signs, pedestrians, other vehicles, etc. The system fuses various observations, makes a decision, and acts locally at the edge (within each car),  including the control of the steering wheel, brake, speed, acceleration, and engine~\cite{kocic2019end}. Meanwhile, vehicles, roads, and surroundings will synchronize their data with each other through data centers and finally coordinate their behaviors. In the background, different models are trained and updated regularly. 

The computations have severe internal and external dependencies. Internally, a self-driving car's current status and actions would impact its subsequent computations and actions through the observe, fuse, act, and coordinate loop. Externally, a self-driving car's status and action would affect the behaviors of the other vehicles.

\item \textbf{Data management and access pattern:}
%Spatio-temporal and stateless data constitute the comprehensive environment information required by automatic driving.
The autonomous driving system relies on a large number of sensors to provide comprehensive environmental information, such as traffic signs, pedestrians, and weather.
%For example, traffic signs indicate autonomous driving the right way, and autonomous driving should pay more attention to pedestrian safety.
The environment information is continuously written for trajectory planning, surrounding object detection, and autonomous control.
The dataset includes LAS binary, float matrix, CSV file, etc~\cite{cordts2015cityscapes, maddern20171, geiger2013vision}.
Moreover, in the morning and evening rush hour, cars often need real-time path planning to avoid traffic congestion. 
The autonomous driving system needs to handle periodic burst accesses.
%Besides, autonomous driving system upload real-time vehicles, roads, and environment information from edges to datacenter for real-time path planning.
%Unlike stateful data, stateless data refers to the sensors' data that is independent of the status of the service object.
%The service object of autonomous driving is humans, but none of the sensors collects human information.
%Stateless data refers to real-time indication information such as traffic signals and petrol gauges.
%Therefore, stateless data guides autonomous driving when the car can go and when it needs to stop.
%Changing road conditions lead to random access patterns for autonomous driving.

\end{itemize}

\subsection{Why we need to build an HFC system?}~\label{2-whybuild}
%% By Gao Wanling & Zhang Wenli %%

%\subsubsection{The Concept of High-fusion Computers}

\subsubsection{The three unique requirements of HFC systems}~\label{HFC_unique_requirement}

%Significantly different from the traditional infrastructures like datacenter, the new class of PSC systems has the following unique requirements. On the one hand, the system should have the ability to perform safety-critical and mission-critical tasks of emerging and future applications, under the constraints of the worst-case, average-case, and best-case performance requriements. Considering the irrevocable effects of actions, the system should accurately verify and validate its decision and collaborate well with the human's decision. 
%On the other hand, the system should support various computation, networking, and data access patterns. 

Significantly different from the warehouse-scale computers that non-stop serve user requests~\cite{GCP2016_google}, HFC systems have three unique requirements. First, they directly interact with the physical world --considering humans as an essential part: human-in-the-loop~\cite{nunes2015survey}, and perform safety-critical and mission-critical operations. Each action may have an irreversible effect. Some systems treat humans as "an external and unpredictable element in the control loop"~\cite{nunes2015survey}, which we call unreliable-human-in-the-loop. For example, many security systems rely on a "human in the loop" to perform security-critical functions~\cite{cranor2008framework}, but humans are incompetent and often fail in their security roles~\cite{cranor2008framework};  In contrast, the other systems consider humans, who make the final decision, a reliable component in the control loop, which we call reliable-human-in-the-loop. For example, medical expert plays a decisive role in medical emergency management systems.
Meanwhile, more scenarios "bolster a closer tie with the human through human-in-the-loop controls that consider human skills, intents, psychological states, emotions, and actions inferred through sensory data"~\cite{nunes2015survey}, which we call collaborative-human-in-the-loop. Collaborative-human-in-the-loop indicates that humans are complements of the other components of the system. Still, an uncoordinated collaboration between a human being and other system components may result in disaster.

%Second, under different conditions like best-case, worst-case, or average-case, the performance, quality and interpretability of computation results are equally critical. 
%For example, as a typical scenario under the worst-case condition, autonomous driving cares about not only the worst-case performance (tail latency) besides the average-case performance (throughput), but also the accuracy of the decision (quality) and why makes this decision (interpretability). 

 Second, unlike Internet services that process independent concurrent requests on planet-scale data center infrastructures, HFC computations have intertwined dependencies between not only adjacent execution loops but also actions or decisions triggered by IoTs, edge, data centers, or humans.  Third, under this extremely entangled state, the systems must first satisfy the accuracy metric in predicting, interpreting, or taking action before meeting the performance goal under different conditions, like worst-case, average-case, and best-case.
 
We take autonomous driving as an example. The systems directly interact with the world. Each action has an irreversible effect. The current status and action of a self-driving car would impact its subsequent computations and actions; a self-driving car's status and action would affect the behaviors of the other vehicles. Autonomous driving must first ensure the accuracy of the decision (quality) and then guarantee the worst-case performance (tail latency). 
%Unlike the planet-scale datacenter infrastructures that mainly serve requests traversed across different data centers, the interaction of PSC systems traverse different paths through and around IoTs, edges, and datacenters and human loops, raising huge requirements and challenges. 

%\subsubsection{HFC has a performance gap of several orders of magnitude  beyond the reach of the state-of-the-practice systems}~\label{performance_gap}
\subsubsection{HFCs demand resources that are several orders of magnitude  beyond the reach of the state-of-the-practice systems}~\label{performance_gap}

Even only taking the worst-case performance metrics -- tail latency -- as examples, we illustrate that the state-of-the-practice systems are far from satisfying the processing requirements.
%From the perspective of quality (e.g., model accuracy) and worst-case performance requirements, which are critical metrics, 
Even for a much simpler application with simpler computation patterns (our motivating example) compared to those in Table~\ref{comparition_table} -- a simplified smart home application with 95\% queries reporting the status and sending heartbeat packets, and 5\% queries processing user requests and accessing the Redis database, serving vast concurrent connections is still tricky.
Zhang et al.~\cite{2020Labeled} simulate this application using a million-level client load generator (MCC) and evaluate the service capacity of kernel TCP and mTCP v2.1  on an X86 server equipped with Intel Xeon E5645 processor, Centos 7.2, Kernel 3.10.0, and 64 GB memory.
%Zhang et al.~\cite{2020Labeled} simulate a smart home application with 95\% queries reporting the status and sending heartbeat packets, and 5\% queries processing user requests and accessing the Redis database. They evaluate the service capacity of kernel TCP and mTCP v2.1 using a client load generator on an X86 server equipped with Intel Xeon E5645 processor, Centos 7.2, Kernel 3.10.0, and 64 GB memory.
Taking 50 milliseconds as the 99th percentile latency threshold,
the kernel TCP supports one hundred thousand concurrent connections, while for the user-level mTCP network stack, the number is nine hundred and sixty thousand~\cite{2020Labeled}.  Accordingly, to achieve ten billion concurrent connections under the threshold of 50 milliseconds, more than ten thousand nodes, even one hundred thousand nodes, would be needed. %Even with the state-of-the-practice system proposed by Zhang et al.~\cite{2020Labeled}, the amount still achieves one thousand. 

Considering the three unique requirements discussed in Section~\ref{HFC_unique_requirement} and the other factors, 
%HFC has a performance gap of several orders of magnitude beyond the reach of the state-of-the-practice systems. 
HFCs demand resources that are several orders of magnitude beyond the reach of the state-of-the-practice systems.
The other factors considered in this rough estimation include scheduling, complex computation and interaction patterns,  complex data access patterns, heterogeneous systems and networks, longer communication links,  and differentiated processing abilities of IoTs, edges and data centers, which will aggravate the situation exponentially.

%Each IoT works within different conditions. Each computation may trigger different algorithms, involve data with different spatial or temporal scopes, and 

%Last, 
%manifesting diverse data access patterns, such as the read and write in real-time, random, burst, periodic, non-periodic, and batch manners.
%IoT devices have a vast number that significantly outweighs the size of the internet users, with notable discrepant functions.

%Considering the much more complex computations and stringent requirements of emerging and future applications compared to traditional applications, we argue that the PSC has a performance gap of several orders of magnitude beyond the reach of the state-of-the-practice systems. 

\begin{figure}[h]
\centering
\includegraphics[scale=0.5]{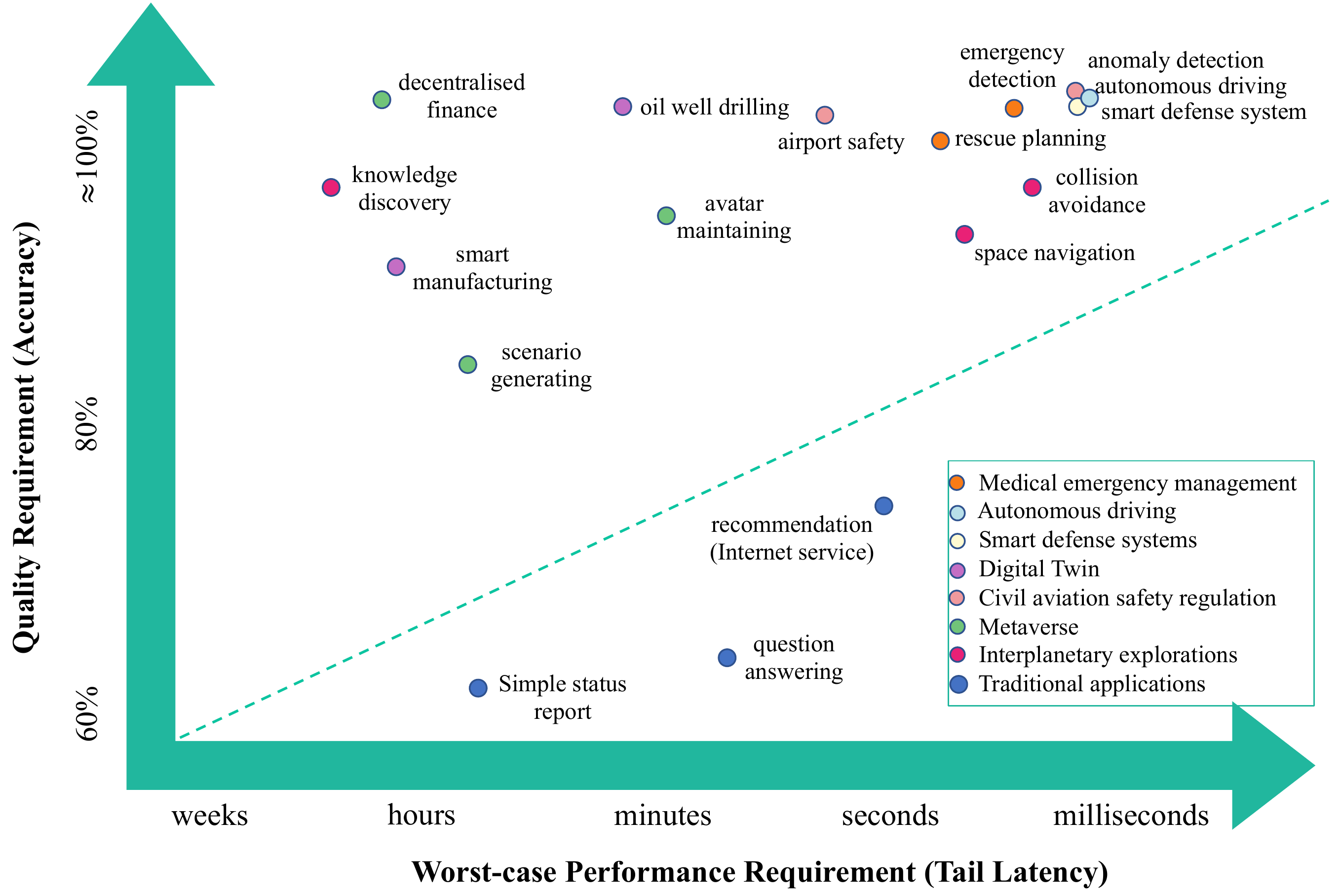}
\caption{This figure compares the worst-case quality and performance requirements between the emerging and future applications in Table 1 and the traditional applications. The quality requirement indicates the required model accuracy of the application or tasks; for example, autonomous driving requires nearly 100\% accuracy to assure safety. The worst-case performance requirement illustrates the required tail latency of the application. For example, autonomous driving requires extremely low tail latency within several milliseconds.}
\label{motivation-gap}
\end{figure}

 We further present several factors in detail. 
  For example, as shown in Fig.~\ref{motivation-gap}, a lot of emerging and future applications usually require nearly perfect quality (i.e., predicting and interpreting accuracy in  Autonomous driving)  and the worst-case tail latency within several milliseconds, which are overlooked in the above motivating example.
  
From the perspective of task scheduling, different scheduling strategies may significantly impact performance. The previous experiments reveal that different placement and scheduling policies of data and workloads across IoTs, edges, and data centers may substantially affect the overall performance: even with the same infrastructure, the gap may achieve dozens or even hundreds of times considering the total response time~\cite{hao2021ai}. 
The scheduling strategies are affected by multiple factors, including task complexity, device processing capability, available resources, network condition, pending tasks, etc. However, the state-of-the-practice solutions provide separate management of IoT, edge, and data center, lacking a global perspective, and further hardly to discover the optimal scheduling strategy. Consequently, unified management and efficient collaboration across IoTs, edges, and data centers are required to assure high performance and resource utilization. 

Instead of simple status reports in our motivating example,  HFC computations are much more complex and intricate, such as object recognition interference, OODA (observe, orient, decide, and act)~\cite{richards2020boyd} or even complex Avatar behavior in terms of machine learning or big data processing; An HFC system manages multi-source and heterogeneous data from various IoT devices, 
manifesting diverse data access patterns in real-time, random, burst, periodic, non-periodic, and batch manners, which also significantly impact the performance;  IoT devices have a vast number that substantially outweighs the size of Internet users, with notable discrepant functions.

\section{The challenges}

This section lays out several HFC challenges.

%in case planet-scale computers are built for emerging and future applications. The general difficulties of open-source computer systems are also discussed in~\cite{ZHAN2022OSSystem}.
% illustrates the details taking Intelligent Transportation application as an example.

\textbf{ Organizability and manageability challenges.} Unlike a traditional computer system, e.g., a supercomputer or warehouse-scale computer, an HFC system is geographically distributed, consisting of IoTs, edges, data centers, and humans-in-the-loop. Moreover, they are dynamic. For example, in smart defense systems and applications~\cite{national2001embedded}, sensors or terminal control units can be dispersed by airdrop, inserted by artillery, and/or individually placed by an operation team~\cite{national2001embedded}. In an extreme case, the spatial realm even has no bound. For example, unmanned spacecraft may have no bounded destination in interplanetary exploration applications. 

So it is challenging to discover the resources and assemble them into a  computer system. The challenges lie in managing these resources efficiently, keeping their survivability in the case of highly possible failures, guaranteeing the immunizability from malicious intrusions and attacks, and improving the programmability and schedulability of massive funclets across a large scale of IoTs, edges, and data centers.

%{\color{blue}{
%\textbf{ Programming challenges.} The PSC systems need to assemble all software and hardware components of IoT, edge, and datacenter as a whole computer. The challenges are how to organize these components as a computer rather than a distributed system, how to manage these resources, and how to find the optimal scheduling strategy. 
%}}

\textbf{Collaboration challenges between software, hardware and people components.} More systems exhibit a closer tie with the human through human-in-the-loop controls~\cite{cranor2008framework}.
However, not all these controls are reliable-human-in-the-loop. For example, some systems treat humans as "an external and unpredictable element in the control loop"~\cite{nunes2015survey}; thus, the action may have an irreversible effect and result in disaster.
The challenge is (1) how do we handle the dilemma of choosing between the system and humans' decision, especially considering "human skills, intents, psychological states, emotions, and actions inferred through sensory data"~\cite{nunes2015survey}?
(2) how do we support the collaboration between the system and humans? (3) how do we decide the respective responsibility in the partnerships? 
On the other hand, when the system makes a solo decision, especially for safety-critical functions or worst-case performance, how do we verify and validate its behavior? What is the human' responsibility behind the system? Let's look at these challenges from the perspective of intelligent defense systems described in Table~\ref{comparition_table}. These challenges are not abstract but vivid and concrete in terms of casualties and losses of people and equipment.

%Each action may have an irreversible effect and hence mandates the interpretability of the HFC system and applications. 

\textbf{Irreversible effect challenges.} Most HFC systems or applications perform safety-critical and mission-critical operations, directly interacting with the physical world. 
Each action may have an irreversible effect under unreliable-human-in-the-loop and collaborative-human-in-the-loop conditions. Even with humans-in-the-loop, considering the human's reaction time, it's hard to make a timely decision and action in an emergency. This irreversible effect demands that the systems' behaviors be verified and validated in advance; the systems can trace the impact of its attributing factors or causes or even achieve interpretability.

 Most HFC systems need to explicitly state the quality of computation results and performance constraints in the entire process, including design, implementation, verification, and validation, referring to its target applications' correctness and performance constraints.   
%The correctness and performance constraints of applications determine their preferred system designs. 
For example, an autonomous driving application requires a worst-case design that assures high accuracy and tail latency within several milliseconds. In this case, we need to holistically verify and validate the systems and algorithms in terms of quality and performance in different cases like best-case, worst-case, and average-case. We have not gained enough experience in this regard.

\textbf{Ecosystem wall challenge}. An entire HFC ecosystem not only consists of the ensemble of the respective IoT, edge, and data center ecosystems but also involves multifarious meanwhile disparate technologies like processor design, operating system, toolchain, middleware, networking, etc.   Moreover,  the design of IoT, edge, and data centers follows distinct guidelines and targets according to their unique requirements or constraints, thus generating various ecosystems with different scopes and boundaries, which we call ecosystem walls. For example, the processor design of IoT pursues low energy consumption and a small chip area, while the data center regards performance as the first element. The daunting complexities caution us that we can not reinvent the wheel. Instead, we need to be compatible with the ecosystem while improving the performance, energy efficiency, and other primary metrics to generate a positive change force that overcomes the ecosystem inertia~\cite{ZHAN2022OSSystem}.

\textbf{The effective evaluation challenges.}
Generally, we need to deploy a system in a real-world environment and run a real-world application scenario to evaluate the performance and provide optimization guidelines. However, the real-world environment and emerging/future application scenarios are inaccessible and costly in assessing and verifying an HFC system. Hence, benchmarks as proxies of emerging/future application scenarios and simulators that emulate real-world systems are necessities for designing, evaluating and optimizing an HFC system.

\begin{itemize}

\item Benchmarking challenges. The real-world emerging or future application scenarios are incredibly complex, involving the interconnection of IoT, edge,  data center, and human-in-the-loop, and including intricate and lengthy execution path~\cite{gao2021aibench}. Constructing benchmarks for HFC systems faces significant challenges. First, the benchmarks should reflect the three unique requirements discussed in Section~\ref{HFC_unique_requirement}, which are not easily embodied in a benchmark manner. Second, the application designers and providers are concerned about the interaction, interconnection, task assignment, and end-to-end performance across IoT, edge, data centers, and human-in-the-loop. Thus, the reality of a benchmark is fundamental. Considering the complexity and confidential issues, the real-world application scenario is not fit to be used directly as a benchmark; hence, a simplified scenario is necessary. However, the real-world scenario contains hundreds or thousands of modules and components; even a tracing or monitoring tool can hardly figure out the execution path and call graph. Simplifying the real-world scenario while maintaining the critical parts is a big challenge. Moreover, considering the humans-in-the-loop behaviors while constructing the benchmarks are a complex problem.

\item Simulation and validation challenges. At the early stage of system and architecture evaluation, the simulator plays a vital role due to the vast manufacturing investment of time and money and the immaturity of the corresponding ecosystem. For example, the effectiveness of the improved processor design, memory access technologies, etc., is evaluated on a simulator. Considering the cost of building a real-world HFC system,  a simulation or validation system supporting the whole environment simulation and technology verification is significant. However, the complexity and diversity of application scenarios pose substantial challenges in building such a simulator.

First, there has no unified interface for different application scenarios or architectures like IoT, edge, and data center. Thus, it is challenging to manage different architectures and support various scenarios. 

Second, simulation accuracy is a crucial metric. High accuracy means the simulator can reflect similar running characteristics to the real world and exhibit running differences under different system environments. Considering the difficulties of multiple-level or multiple-scale simulation, including hardware level, e.g., processor chip, cache, memory hierarchy, disk, and software level, e.g., operation system, ensuring the simulation accuracy is necessary but challenging. 

\end{itemize}

%\subsection{RISC-V}
%\subsection{Key components}

%\textbf{architecture}

%\subsection{Benchmarks}

%\subsubsection{CPU Benchmarks}

%\subsubsection{IoT Benchmarks}

%\clearpage

%\input{section/goals}

%\input{section/standards}

\section{Exploring the solution space}

%\subsection{Memory system challenges}
%% By Chen Mingyu %%

%\subsection{Smart network interfaces} 

%\subsection{Performance-deterministic distributed storage system}

%\subsection{Lightweight network protocol stack}
%% By Zhang Wenli %%

%\subsection{Full-stack cooperative optimization}

%\subsection{Collaborative resource management}
%\subsection{Energy efficiency}

%\subsection{Methodology, reference architecture and open standards}

We adopt a funclet methodology and architecture to facilitate exploring the HFC software and hardware design space. According to~\cite{ZHAN2022OSSystem}, the funclet represents the common proprieties of basic building blocks across the systems. Each funclet has the following characteristics~\cite{ZHAN2022OSSystem}:"each contains a well-defined and evolvable functionality with modest complexity; each can be reusable in different contexts;  each can be independently tested and verified before integrating; each can be independently deployable; each can interoperate with other funclets through a well-defined bus interface or interconnection". 

The funclet architecture consists of four layers: chiplet, HWlet, envlet, and servlet. 
A chiplet is "an integrated circuit (IC) with modest complexity, providing well-defined functionality" ~\cite{li2020chiplet,chipletgoodorbad2020}; it is "designed to be susceptible to integration with other chiplets, connected with a die-to-die interconnect scheme" ~\cite{li2020chiplet,chipletgoodorbad2020}.  A servlet is "an independently deployable and evolvable component that serves users with a well-defined and modest-complexity functionality"~\cite{ZHAN2022OSSystem}. Microservices~\cite{nadareishvili2016microservice} or cloud functions~\cite{jonas2019cloud} are two forms of servlet. An HWlet is "an independently deployable, replaceable, and accessible hardware component, e.g., CPU, memory, storage" ~\cite{ZHAN2022OSSystem}. An envlet is "an independently deployable and evolvable environment component with well-defined functionality that supports the management of servlets"~\cite{ZHAN2022OSSystem} .

The funclet architecture uses a three-tuple: \{funclet set architecture (FSA), organization, system specifics\}~\cite{ZHAN2022OSSystem} methodology to describe the architecture. According to ~\cite{ZHAN2022OSSystem}, the FSA refers to "the actual programmer-visible function set~\cite{hennessy2019computer}, serving as the boundary between two adjacent layers and among different funclets in the same layer"; The organization includes "the high-level aspects of how funclets in the same layer and adjacent layers collaborate";  The system specifics describe "the design and implementation of a system built from funclets" ~\cite{ZHAN2022OSSystem}. 
The advantages of the funclet architecture are discussed in~\cite{ZHAN2022OSSystem}: it increases technology openness, improves productivity, and lowers cost; relieves the complexity of building systems; improves reusability and reliability.

\begin{figure}[h]
\centering
\includegraphics[scale=0.3]{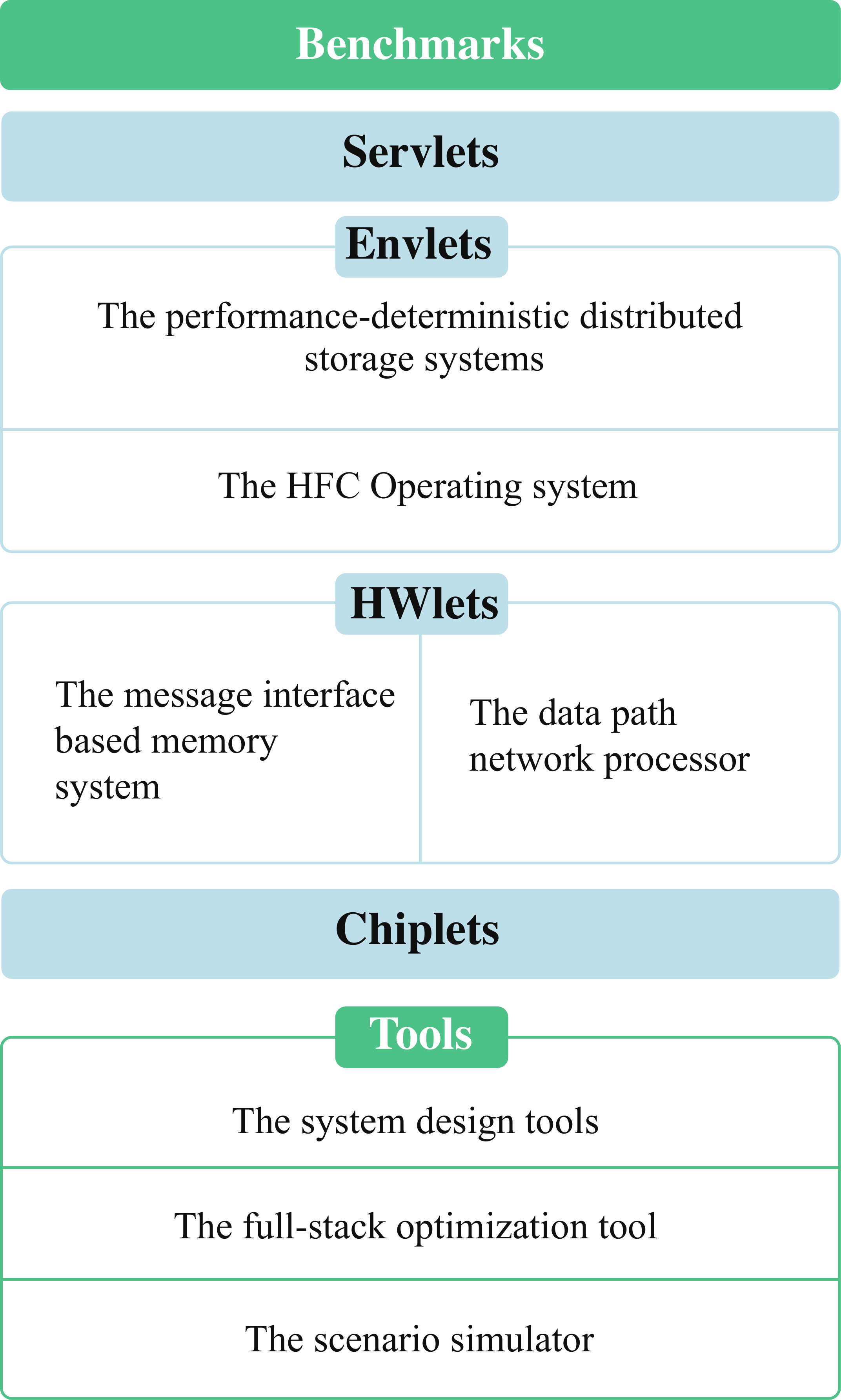}
\caption{The overview of the reference architecture.}
\label{OverviewArch}
\end{figure}

Fig.~\ref{OverviewArch} illustrates the reference HFC architecture and components. For the chiplet layer, we focus on workload-driven chiplet designs. 
%For exploring and validating the design of chiplets, we provide a  processor development tool suite.
For the HWlet layer, we pay attention to the message interface-based memory system and data path network processor. For the envlet layer, we concentrate on the HFC operating systems and the performance-deterministic distributed storage systems. Also, we provide several tools. A system design tool suite is provided to help designers explore the HFC design space across chiplet, HWlet, envlet, and servlet. A scenario simulator is built across IoTs, edges, and data centers to accelerate innovative technologies' deployment and verification;  A full-stack optimization tool is provided.  We construct a series of benchmarks and microservices for various HFC applications for the servlet layer.  Several scenario benchmarks are proposed as the proxy of real-world application scenarios, aiming to support the whole-stack evaluation and provide feedback to scenario simulators or even real-world scenarios. 

%\subsection{Open standards}

\begin{figure*}
	\centering
		\includegraphics[scale=.34]{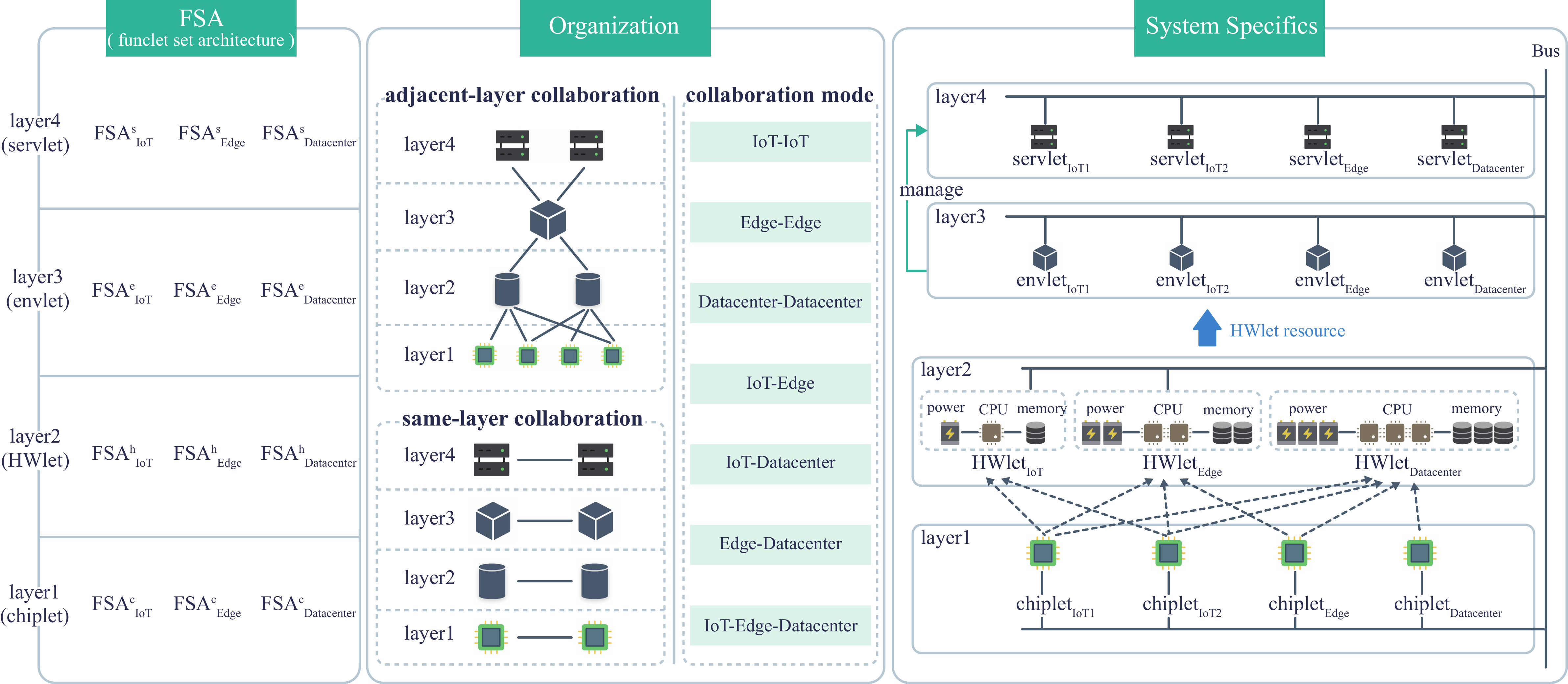}
	\caption{An HFC instance of the four-layer funclet architecture based on~\cite{ZHAN2022OSSystem}. }
	\label{4lets_architecture}
\end{figure*}

Figure~\ref{4lets_architecture} shows a funclet-based HFC architecture in terms of \{funclet set architecture (FSA), organization, system specifics\}, which is derived from an open-source computer system initiative~\cite{ZHAN2022OSSystem}.
%The funclet-based PSC architecture contains four layers: chiplet, HWlet, envlet, and servlet, which defines the integrated circuit with well-defined functionality, hardware components, environment component, and user services, respectively. 
Each layer of chiplet, HWlet, envlet, and servlet specifies a set of \emph{FSAs} for IoTs, edges, and data centers. The \emph{FSAs} are composable and collaborative through the same-layer and adjacent layer. The organization specifies "the high-level aspects of how funclets in the same layer and adjacent layers collaborate"~\cite{ZHAN2022OSSystem}, including IoT-IoT, Edge-Edge, Datacenter-Datacenter, IoT-Edge, IoT-Datacenter, Edge-Datacenter, and IoT-Edge-Datacenter collaborations. The \emph{system specifies} shows how to implement an HFC system. The different layers for IoTs, edges and data centers are interconnected through the funclet-based open standards, including interconnection, interface, protocol, and networking across IoTs, edges, data centers, and humans-in-the-loop.

\subsection{Benchmarks}

%This subsection presents the benchmarks and applications.

%\subsubsection{Benchmarks}

As the foundation of system design and optimization, benchmarks are of great significance for developing HFC systems. We aim to propose a series of benchmarks that reflect three unique characteristics and other important factors of HFC computations for designing and evaluating HFC systems.
%to cover the measuring, evaluating, and comparing top-level systems, middle-level modules and components, and low-level architectures.

\textbf{Scenario benchmarks}. From the perspective of the whole-stack system benchmarking, e.g., the interconnection and communication of IoTs, edges, data centers, and humans-in-the-loop, we propose scenario benchmarks as the proxy of the real-world application scenario. The construction follows a scenario-distilling methodology that formalizes a real-world application scenario as a graph model and distills it into a combination of essential tasks and components~\cite{gao2021aibench}. This methodology identifies "the critical path and primary modules of a real-world scenario since they consume the most system resources and are the core focuses for system design and optimization"~\cite{gao2021aibench}, thus reducing complexity.

\textbf{IoT Benchmarks}. From the perspective of middle-level modular benchmarking, we construct IoT benchmarks to evaluate mobile and embedded devices. The IoT benchmarks include the lightweight IoT workloads and light-weight AI workloads.

\textbf{CPU Benchmarks}. We present CPU benchmarks covering typical workloads  from emerging and future application scenarios from the low-level architecture benchmarking perspective. Constructing CPU benchmarks adopts a traceable methodology, managing the traceable processes from problem definition, problem instantiation, solution instantiation, and measurement~\cite{zhan2022benchcouncil}.
%whole-picture workload characterization (WPC) 
%methodology~\cite{wang2022wpc} and comprehensively analyzes the intermediate representation (IR) level, ISA level, and microarchitecture level performance. 

%\subsubsubsection{Scenario Benchmarks}

%We propose a scenario-distilling methodology and a reusing framework to construct scenario benchmarks, which formalizes a real-world application scenario as a graph model and distills it into a permutation of essential tasks and components~\cite{gao2021aibench}. This methodology preserves the critical path and essentially reduces the complexity. Our previous work adopts a gray-box scenario-distilling methodology that accesses no code but needs the industry's design input. It models the real-world one as a Directed Acyclic Graph-based model and provides two scenario benchmarks---E-commerce Search Intelligence and Online Translation Intelligence~\cite{gao2021aibench}. 

%Focusing on the novel IoT-edge-datacenter architecture, we aim to provide scenario benchmarks across IoT ends, edge computing, and datacenter. It needs to reflect the interaction and interconnection among these three dimensions. 
%In addition, considering different users and different benchmarking requirements, we are also dedicated to a white-box (apply to industry users who can access the code) and a black-box (without input) methodology. 

%\subsubsection{Applications}

\subsection{Chiplets}

This subsection presents the workload-driven chiplet design methodology to explore the ideal architecture for each class of emerging and future applications. We aim to provide reusable building blocks  considering the different PPA (performance, power, area) requirements for IoT, edge, and data center.

%For the first step, we characterize the computation, memory access and communication patterns of most real-world PSC application scenarios and classify them into several classes of workloads, and each class represents consistent requirements. 

For the first step, we perform comprehensive workload characterization on a broad spectrum of tasks within target applications. For each application in Table~\ref{comparition_table}, we analyze the computation patterns drilling down into the critical computations within the execution loop,  like OODA in smart defense scenarios across IoT, edge, and data center. Then we analyze the interaction patterns covering the interactions between IoT-IoT, IoT-edge, IoT-datacenter, edge-edge, edge-datacenter, and datacenter-datacenter. 
Above all, the analysis contains computation, memory access, networking, and other characteristics.
According to the results, on a single level of IoT, edge, or data center, we classify their characteristics into several classes for each pattern, like computation. After that, we will obtain several classes, including (IoT, computation), (IoT, memory access), (IoT, networking), (edge, computation), (edge, memory access), (edge, networking), (data center, computation), (data center, memory access), (data center, networking), etc.

%For the first step, we perform comprehensive workload characterization on a broad spectrum of tasks within target applications. The analysis contains the computation, memory access, networking, and other patterns covering IoT, edge, and datacenter.
%According to the results, on a single level of IoT, edge, or datacenter, we classify their characteristics into several classes for each pattern like computation. After that, we will obtain several classes including (IoT, computation), (IoT, memory access), (IoT, networking), (edge, computation), (edge, memory access), (edge, networking), (datacenter, computation), (datacenter, memory access), (datacenter, networking), etc. 

%From the perspective of module linking, we classify several critical linkings spanning computation, memory access, networking, etc. After that, we will obtain several linkings like (computation, memory access), (computation, networking), etc.
%From the perspective of adjacent-layer collaboration of IoT, edge, datacenter, we classify the IoT/edge/datacenter interaction patterns into several mainstream combinations. After that, we will obtain several classes including (IoT, edge, computation), (IoT, edge, memory access), (IoT, edge, networking), (edge, datacenter, computation), etc.

For the second step, we attempt to define the ideal chiplet architecture for different IoT, edge, data center layers and different analyzed patterns. Specifically, according to the classifications in Step 1, we define the computation, memory, networking chiplets for IoT, edge, and data center, respectively. Each layer of IoT, edge, or data center will contain multiple chiplets for different patterns of computation, memory access, networking, etc. Additionally, each pattern may contain multiple chiplet designs according to the classifications of workload characteristics.

%For the third step, we explore the packaging technologies to integrate different chiplets according to the linking classifications and adjacent-layer combinations in Step 1. Using the linking classifications, we define the packaging technologies of computation, memory, networking, and other chiplets, including (computation chiplet1, computation chiplet1), (computation chiplet1, computation chplet2), (computation chiplet1, memory chiplet1), (computation chiplet1, memory chiplet2), (memory chiplet1, memory chiplet2), etc. 
%Using the adjacent-layer combinations, we define the packaging technologies of chiplets for IoT, edge, and datacenter, including (IoT computation chiplet1, edge computation chiplet1), (IoT memory chiplet1, edge memory chiplet1), (IoT computation chiplet1, edge memory chiplet1), etc.

For the third step, we validate the chiplet architecture design and further performs improvements according to the feedback. 
%We perform FPGA-based and RTL-based simulation and validation to achieve simulation performance and debuggability. 
%We evaluate the scenario, IoT, and CPU benchmarks to conduct the functionality and performance validation. 
We adopt FPGA-based simulation and evaluate the scenario, IoT, and CPU benchmarks to conduct the functionality and performance validation. 
Further, we explore the upgrades and design optimizations based on the validation results.

The chiplet architecture design contains a loop of workload characterization, chiplet design, and validation until the output designs satisfy the application requirements.

%For example, we single out the critical workloads from different scenarios, like lane detection and traffic light detection from autonomous driving, and  thoroughly analyze their characteristics, and conduct classification.

%For the second step, we attempt to define the ideal chiplet architecture for each class of workloads, considering both the data access and computation schemes. Moreover, the design is also affected by the unique requirements on performance, energy consumption, and chip area. The architecture exploration is based on open-source simulators like GEM5~\cite{binkert2011gem5} or FPGA technology.

%The third step explores the interconnection, interface, protocol, and networking standards among chiplet architectures. By organizing several chiplet architecture classes, we aim to achieve the functionality and performance guarantee at the chiplet layer of a real-world application scenario.

%The fourth step validates the chiplet architecture design and organization and further performs improvements according to the feedback. We evaluate the scenario, IoT, and CPU benchmarks to conduct the functionality and performance validation. Further, we explore the upgrades and design space based on the benchmarking results.

%The chiplet architecture design is a loop of design exploration, benchmarking, and validation until the design satisfies the application requirements.

\subsection{HWlets}
This subsection presents the HWlets solutions, primarily focusing on two innovative HWlets: a message interface-based memory system and a data path network processor.

\subsubsection{The message interface based memory system}

 % Traditionally, the memory system consists of two types -- internal memory and external memory. The internal memory or main memory usually refers to DRAM-based memory systems accessed by a byte-level load/store CPU instructions. The external memory or storage refers to disk-based memory systems accessed by block-level read or /write I/O operations. There is an increasingly obvious trend that the two types of memory will be merged or unified in the future.  
   
%   On the one hand, different non-volatile memory (NVM) technologies have emerged as supplements to dominant capacitor-based DRAM, including PCM, RRAM, MRAM, etc. Second, remote memory access technologies and high-speed network technologies are developing fast. RDMA has been deployed widely. Several coherence extension interfaces, such as CAPI, CCIX, and CXL, compete for industry standards. NVM and remote memory are sometimes called far-memory since they have longer access latency than DRAM. 
   
 %  On the other hand, a solid-state disk dramatically shortens the access latency of storage. Fast-NAND devices and PCM-based storage have entered the range of ten microseconds. The bandwidth of the new storage device is also comparable with the high-speed network. 
   
  % Above all, 
   As the boundary between internal and external memory is blurred, a computer system may face different memory devices with various latency, bandwidth, granularity, and capacity. 
   Thus, the challenge is providing a universal memory interface and a unified memory system so that the programmers do not need to switch between byte-level load/store CPU instructions (for internal memory) or block-level read/write I/O operations. 
   
   %load/store and read/write confusedly and further run the program with a smooth resource distribution among complex memory hierarchy.

%% By Chen Mingyu %%
    
    We have proposed a message-interface-based memory access approach to solve various "memory wall" problems~\cite{MIMS2014}. We plan to extend the message interface to include internal and external memory to build a unified memory system.     The system assumes a high bandwidth, high concurrency, and low latency network, which we believe will come soon.
    
    Instead of only using a fixed command format and address for a memory request, we propose to use a message that contains rich semantics to express a memory request. The semantic information consists of size, sequence, priority, process id, persistence, etc., or even array, link pointer, and locks. Furthermore, the memory resource provider is no longer a simple dumb device but with different local computation capabilities to service a message request.
    
   The message interface base memory system decouples the data access from the data organization. A client does not need to know the details or memory resource organization like banks and rows of an SDRAM, even the exact location of the data. The message interface-based memory system also decouples the data access from data transfer. Small data requests can be combined into a large network message. A large data request can be divided into multiple messages.
    
    There are three critical components to implementing a message interface-based memory system. (1) a CPU core generates concurrent memory requests with semantic information. Traditional load/store-based instructions are too simple to express a rich-semantic memory request. Instead, we need kinds of asynchronous and operand-variable instructions. (2) a memory controller with a message interface. The controller should group and assemble memory requests from internal CPU cores into coarse-grained messages to exchange with different memory servers through the network. (3) various message-interfaced memory servers. The memory server manages local storage media and accepts message requests and responses after specific local processing.  We will implement the previous two components as chiplets and the message-interfaced memory servers as HWlets.

\subsubsection{The data path network processor}    

%% By Chen Mingyu  Hardware acceleration for network stack processing has been studied for a long time.%%
    
    High-speed networks have been an indispensable part of modern computer systems. There are two distinct technical routes for network acceleration: offloading~\cite{liu2019offloading,le2017uno,moon2020acceltcp} and onloading~\cite{dosanjh2015re}. Offloading means to offload part of the network processing to an external accelerator card, namely smart NICs~\cite{yan2020p4,liu2019offloading}, and save CPU resources for application logic. For example, the checksum of TCP packets can be computed on high-end NICs~\cite{moon2020acceltcp}. Although various smart NICs have been proposed, none of them has occupied the dominant position in the market. More recently, smart NICs
     have a new name -- DPU (Data processing unit)~\cite{burstein2021nvidia}. On the other hand, onloading means pushing all network packet processing onto general-purpose CPU cores to utilize the ever-increasing computing power of the CPU core. DPDK by Intel~\cite{dpdk} is the industry standard for onloading with the help of hardware features built-in Intel CPUs.

    Whether offloading or onloading, there are many functions in a network stack that can be accelerated by hardware, for example, checksum, encryption/decryption, table lookup, keyword matching, queuing, ordering, etc. Additionally, several functions can only be efficiently processed by the CPU, such as fragments, lists, buffers, order,  and other complex data structures.
    
    Hence, we argue that the critical point is organizing these accelerating resources efficiently. Equipping a general-purpose CPU with accelerators accessed via the I/O bus is not the most efficient solution. Likewise, putting a powerful general-purpose multi-core CPU into a NIC will not change much.     The challenge here is how to design efficient control and data path of accelerators inside a CPU to combine both the accelerator units' special functions and the CPU cores' general processing ability. 

%% By Chen Mingyu %%
    For general-purpose processors, NIC is always an "external" device. The processor has to initialize a DMA operation to move data from the I/O bus to the local memory or cache to access packet data. Most built-in processors in smart NICs still have such structures. That results in uncontrollable processing latency, so these processors are only suitable for processing control paths. Only hardware logic like FPGA or special function-limited core like P4 engine can process line-speed data paths. They can reach the line speed only because their functions are simple and deterministic enough. Generally, they cannot process complex semantic information in data paths that need complex data structures to store the state of many concurrent transactions.

    We propose to design a processor for line-speed processing data paths, which we call the data path network processor (a datapath processor). We will implement the datapath processor as an accelerator. However, the network packet will be the first-class citizen in the datapath processor. The packet stream leaves the register file of the primary CPU and arrives at the datapath processor directly without going through a complex memory hierarchy. The datapath processor has full functionality, including access to the cache and main memory. Thus, the datapath processor can hold and process complex state information necessary for the data plane. The datapath processor also has accelerating units on the local bus; data exchange between the datapath processor core and accelerator can be low latency, highly paralleled, and fine-grain.     We have not seen such the structure of the datapath processor before, but fortunately, open-source processors, like RISC-V based, allow us to design a novel processor architecture freely. 

\subsection{Envlets}

This subsection presents the Envlets solutions, primarily focusing on the HFC OS and the HFC distributed storage systems. %, and collaborative resource management. 

\subsubsection{The HFC operating system}
%% by Chen Zheng %%
%With the rapid development of emerging applications and hardware, modern computing scenarios have quickly turned into a discrete pattern but with more fusion needs. Computing is ubiquitous but needs a more efficient way of organizing. As the basis software stack of computing systems, a brand new operating system (OS) model that integrates and adapts to the IoT-edge-datacenter architecture is emerging and has become a hot spot for system software innovation and industry interest. OS needs to manage and drive heterogeneous devices in a real environment, including various sensors, actuators, traditional processors, intelligent accelerators, industrial robots, datacenter servers, etc.

%The hardware devices are heterogeneous, distributed ubiquitously, and with different performance and power consumption. The hardware management should meet high reliability, security, real-time performance, and access efficiency requirements. These emerging computing requirements lead us to build a new OS architecture to drive the heterogeneous devices' hardware resources, interconnect distributed devices, provide optimized OS run-time for different workloads, serve workloads with a more intelligent scheduling policy, and solve the issues in multi-device collaboration. Building the new OS architecture faces the following challenges:
%% Revised by Wang Lei%%

In the HFC scenarios, computing is ubiquitous, consisting of geographically distributed, heterogeneous hardware devices with different performance and power consumption constraints. Hence, we need a more efficient way to improve the organizability and manageability of the HFC systems. Our OS solutions aim to provide the following features: 

(1) The new OS should have a flexible system structure. For different devices and workloads, specific OS capabilities need to be built. OS is no longer limited to a single kernel running directly in the local node but a distributed OS architecture that adapts to hardware resources in different computing nodes. OS needs to be able to reconfigure or rebuild itself in run-time to adapt to various scenarios.

(2) To efficiently manage the distributed heterogeneous hardware, the new OS should rebuild a general and intelligent device driving framework to discover, identify, register, access, and drive the massive hardware resources automatically. In addition, it can establish a soft bus connection for interactions with the immunizability  from the malicious intrusions and attacks.

(3) To meet different HFC workloads' performance targets, the new OS needs to build fine-grained resource metering and application profiling features to facilitate efficient scheduling for improving performance and resource utilization.

(4) The HFC ecological boundary is open. As a result, OS faces security challenges, such as  end-to-end device authentication access, identification issues in an open environment, and security isolation. Under the premise of "zero trust," we need to embody security enhancement strategies in the native OS kernel.

%The emerging computing ecosystem includes typical scenarios such as autonomous driving, medical emergency management, and industrial Internet. The workloads are complex, and the hardware is distributed and heterogeneous. As a result, The new OS should have a flexible system structure. %It is urgent to establish the basic OS theory and architecture research to propose the OS construction method to lay the foundation for designing and implementing the new operating system. 
%To efficiently manage the distributed heterogeneous hardware, %traditional OS mainly faces two issues: it is difficult for OS to discover and identify the hardware efficiently; it is difficult to synchronize and interact across multi-devices effectively. 
%Traditional OS cannot guarantee the execution efficiency of AI algorithms from the system level as the lack of hardware capability knowledge and intelligent scheduling policies. 

\subsubsection{The performance-deterministic distributed storage systems}

%% By Xiong Jin %%

%% By Xiong Jin %%

The emerging and future applications %like autonomous driving listed in Table~\ref{} 
heavily rely on advanced techniques like big data and AI, performing hybrid and concurrent tasks with different requirements, e.g., latency-critical and throughput-critical, and pursuing the worst, average, or best-case performance. However, these hybrid tasks usually adopt different systems and architectures, and have distinct data access patterns and requirements~\cite{bigdata2014,dlio2018}. In addition, the worst-case tail latency poses great challenges even for Internet services in data centers~\cite{tail2013,qoe2001,qoe2012}, let alone much more complex applications across IoTs, edges, and data centers. Thus, to efficiently serve these applications and tasks, building a single distributed storage system (DSS) that provides
\textit{deterministic performance and high throughput} is an urgent demand~\cite{cake2012, reflex2017, qwin2021}. Note that the deterministic performance means that a DSS should enforce differentiated tail latency SLOs for concurrent latency-critical tasks. Throughput means the total QPS (requests per second) or bandwidth of a DSS.

The design of a DSS needs to consider the characteristics of storage devices. We conclude two development trends of storage devices. On the one hand, the devices will be increasingly faster with microsecond-scale or even lower latency. 
Storage devices have experienced several technological breakthroughs in the past twenty years, such as the development of commercial SSD products, NVM-based SSD products (Intel Optane SSD~\cite{iossd}), and persistent memory products (Intel Optane PM~\cite{iopm}). Compared to HDD and ordinary SSD, NVM-base devices have much lower latency. In addition, emerging fast networks (e.g., 200 Gbps and 400 Gbps Infiniband) have round-trip latency of less than 1 $\mu$s~\cite{infiniband2021}. These low latency devices put forward high demand to the storage systems~\cite{killer2017}. 
On the other hand, the devices will contain enhanced computation capacities, such as computational storage drives~\cite{smartssd,csd}, SmarkNICs~\cite{smartnic}, and programmable switches~\cite{pswitch}. Many studies propose to offload several tasks to the devices and have shown the performance advantages, like offloading query processing and data (de-)compression to SmartSSDs~\cite{query2013,biscuit2016,ysql2016,cidr2019}, data replication and file system functions to SmartNIC~\cite{hloop2018,linefs2021}, and global memory management, load balance and data cache to programmable switches~\cite{mind2021,pegasus2020,dcache2019}. 
Hence, the design of a DSS needs to make full use of these in-device computing resources.

Considering the application requirements and device characteristics, building a DSS faces serious challenges. First, due to the distinct states of different machines/threads, latency spikes~\cite{tail2013,tail2014}, schedulability issues~\cite{sched2018}, load burst~\cite{arachne2018,shenango2019,shinjuku2019,perfiso2018}, and resource contention inside storage devices and the network stack~\cite{tails2016,ttail2017,llssd2019}, it is extremely hard to guarantee deterministic performance and high throughput. %Many factors such as CPU scheduling, background system routines, garbage collection of memory space, and lock contention in the software could bring latency spikes. 
%Second, the unschedulable points in the I/O path of the current DSS result in schedulability issues, and further bring in difficulties for providing deterministic performance~\cite{sched2018}.
%Third, resource contention inside storage devices and the network stack may cause severe performance variations, especially latency. The gap may even achieve to hundreds of thousands of times~\cite{tails2016,ttail2017,llssd2019}.
Second, many technologies have been proposed to achieve low latency, including using poll instead of interrupts~\cite{poll2012}, kernel-bypass I/O like user-space communication mechanism~\cite{rdma2007,dpdk}, user-space device drivers~\cite{nvmed2016,spdk2017}, and user-space file systems~\cite{fsp2019,ufs2021}. However, the previous DSS only adopt a single technology, and it is challenging to integrate and benefit from all these technologies in a single DSS.  
Third, there are increasing computation capacities inside devices through either FPGA or low-energy embedded processors. Previous work attempts to offload partial computation to in-device computing logic~\cite{nddl2019,dnn2019,aics2020,mlt2021,dnn2021}. We argue that offloading partial infrastructure software like DSS~\cite{linefs2021,fusionfs2022} and SQL engine~\cite{query2013,ysql2016} rather than user applications are better~\cite{linefs2021}. However, due to the complicated functionalities of DSS, it is a tough thing.

Our solutions for a novel DSS include the following innovations. 
First, a new DSS should exploit the leading technologies for emerging devices with $\mu$s-scale latencies, including polling device events, user-space I/O, and run-to-completion request processing. Existing efforts focus on one individual technology and fail to integrate all of them into a single system in a systematic way for efficiency. %For example, eliminate possible overhead along the I/O path such as data copies and CPU wastes. 
Second, a new DSS should embed scheduling along the whole I/O path from the clients to the servers and storage devices. Schedulable architectures should be used at each layer along the I/O path, including client-side, network, server-side, and storage devices. Moreover, exploiting in-device computing power through software-hardware co-design controls resource utilization at the fine granularity and reduces latency. 
Third, a new DSS should offload some functionalities to the devices without reducing total throughput and impairing the performance of individual applications. Existing efforts exploit one type of in-device computing power, either computational storage, SmartNIC, or programmable switches. Unlike them, a new DSS should make full use of all these in-device computing powers to achieve better performance and energy efficiency.

\subsection{Tools}

This section presents our tools: the system design tools, the full-stack optimization tool, and the scenario simulator.

\subsubsection{The system design tools}
We propose system design tools to help designers explore the HFC design space. The system design tools include the chiplet design tool, the HWlet design tool, the envlet design tool, and the servlet design tool, which correspond to the funclet architecture. Each design tool provides the simulation-validation-development tool suite. For example, for the chiplet design tool, we propose a whole-picture simulation to explore the co-design space of the chiplet across stacks. Unlike the traditional microarchitecture-level simulation, such as the GEM5, our whole-picture simulation is across full system stack levels, including three hierarchical levels: the IR (intermediate representation ) level, ISA, and microarchitecture levels~\cite{wang2022wpc}. The whole-picture simulation combines the IR, ISA, and microarchitecture level simulations. The design decisions at IR, ISA, and microarchitecture levels are ISA-independent, microarchitecture-independent, or specific to the actual processor's microarchitecture. Combining the design decisions from the IR, ISA, and microarchitecture, the user can explore the co-design space across stacks.
Furthermore, we propose cycle-accurate and bit-accurate circuit simulations and verification tools for the validation. The validation tool is based on general x86 computing and heterogeneous FPGA resources and provides an on-demand service for chiplet validation. For the development,  we propose AI-based open-sourced EDA tools for integrated circuit design and development.

\subsubsection{The full-stack optimization tool}

%% By Luo Chunjie %%
The full-stack optimization tool has the following challenges:

(1) The optimization object is uncertain. Finding the performance bottleneck in an HFC system is non-trivial. Users may feel confused about the optimization objects because of the optimization possibilities on IoTs, edges, or data centers and the complex hierarchies of algorithms, frameworks, software, and hardware. %Also, they may be uncertainty optimized on the IoTs, edges, or clouds. The discovery of bottlenecks requires experienced full-stack engineers.

(2) The optimization space is vast. There are thousands of optimization dimensions of the algorithm, software, and hardware, and the values of the variate vary in an extensive range. As a result, the optimization space is exceptionally huge. 

(3) The optimization target is diverse. Different application scenarios have additional user requirements. In addition to the vital importance of accuracy, some applications are sensitive to latency; some require high throughput, and some are concerned with energy consumption. Therefore, different application scenarios have other optimization goals.

%% By Luo Chunjie %%
The tool covers the optimization from vertical and horizontal dimensions. From the vertical dimension,
we will co-explore the optimization space from the algorithm, software, and hardware. For example, for deep learning applications,  jointly optimizing the network's architecture and the hardware accelerators is promising in improving performance and reducing energy consumption. We will consider the close collaboration among IoTs, Edges, data centers, and humans-in-the-loop from the horizontal dimension. For example, we will automatically offload whole or partial deep neural network computations from end devices to more powerful devices, such as edges or data centers. 

Automatically co-optimization is non-trivial because of the vast optimization space. There are thousands of optimization dimensions of the algorithm, software, and hardware, and the values of the variate vary in an extensive range. As a result, the optimization space is too huge to complete the search. Reinforcement learning has shown powerful capabilities for the problem of searching for optimal policies in a vast space. Evaluating is expensive in co-optimizing the algorithm, software, and hardware across the IoTs, edges, and data center. We will investigate the state-of-the-art learning algorithms and evaluation strategies and develop the corresponding tools for automatic optimization.

\subsubsection{The scenario simulator}
The scenario simulator is a miniature of the real system, which contains unified interfaces and replaceable components to enable rapid deployment and verification of innovative technologies. The scenario simulator manages the whole environment of a computer system, e.g., processor chip, operating system, memory, network, etc. It covers the complete execution and interaction across IoTs, edges, data centers, and humans-in-the-loop. It can demonstrate the effects visually, e.g., running results, performance, and power consumption, under different technologies, deployments, or parameter settings, for example, the latency performance of an autodrive scenario using other memory devices and network protocols. For the first step, we plan to provide a network simulator that simulates the communication patterns of representative scenarios like big data and artificial intelligence, supporting different networking technologies. The involved components are replaceable, and we can easily use emerging technology to replace the existing one and verify its effectiveness. We will expand the scenario simulator to the whole HFC environment for the next step.

\section{Our plan}

We aim to define a new paradigm -- IoTs, edges, data centers, and humans-in-the-loop as a computer and launch an open-source high fusion computer (HFC) system initiative. The goal is to vastly enhance the system capabilities under specific energy and cost constraints for most emerging and future applications.

We abstract reusable functions (funclets) across system stacks among IoTs, edges, and data centers to guide the HFC system design and evaluation. We first propose to define a series of benchmarks and funclet-based standards and then build the tools to facilitate the workload-driven exploration of the system and architecture design space.
%After that, we provide open-source tools on EDA and FPGA solutions. 
Finally, we provide open-source implementations of an HFC system. We will perform system co-design from the vertical and horizontal dimensions throughout the process. Vertically, we comprehensively explore the algorithms, runtime systems, resource management, storage, memory, networking, and chip technologies. Horizontally, we deeply discover the collaboration and interaction among IoT, edges, data centers, and humans-in-the-loop.

\begin{figure*}
	\centering
		\includegraphics[scale=0.52]{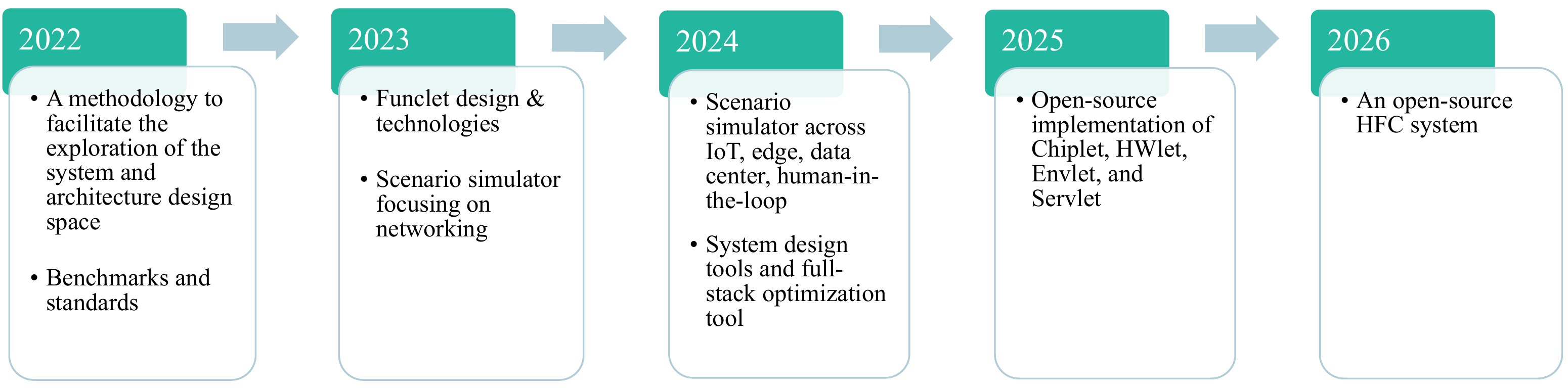}
	\caption{Development milestone of HFC system.}
	\label{5plan}
\end{figure*}

We plan to build the first open-source implementation of an HFC system using an iterative and evolving way. Fig.~\ref{5plan} shows the development milestone of our HFC system. We first focus on one or two typical application scenarios and essential funclets across IoT, edge, and data centers to reduce the complexity. Then we expand the focus and update or replace the technologies gradually. A scenario simulator is beneficial to the expansion, update, and replacement. Finally, we will summarize the experience and lessons during this period and dedicate ourselves to the contributions of a useful HFC system and related advanced technologies. 

\section{Related work}

\begin{figure*}
	\centering
		\includegraphics[scale=0.4]{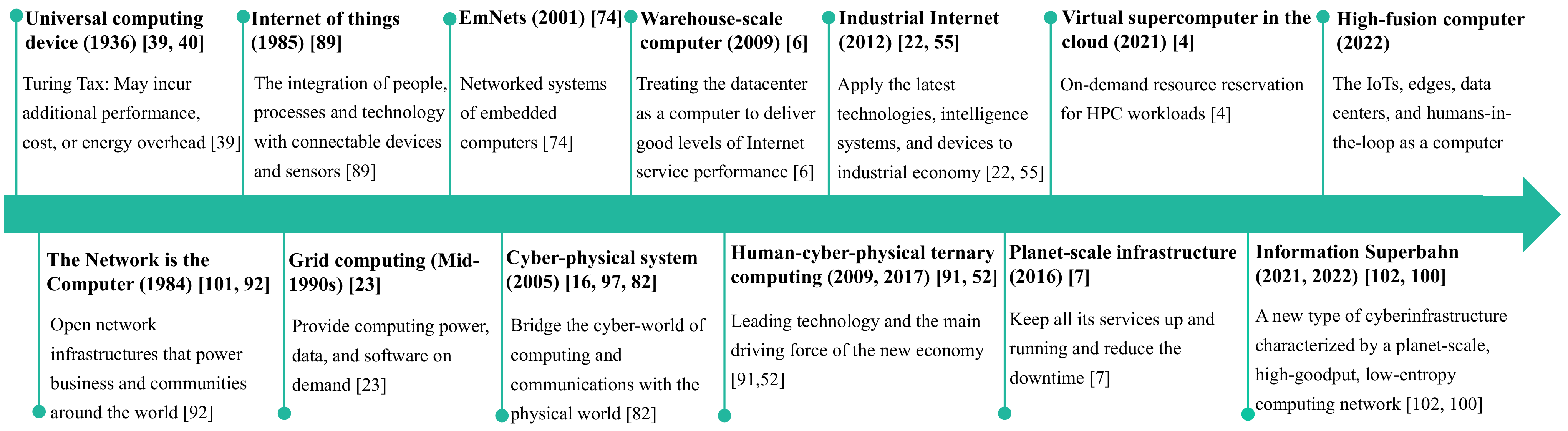}
	\caption{An overview of the related work.}
	\label{6relate}
\end{figure*}

The development of the computer industry witnessed a series of computer systems concepts and implementations, as shown in Fig.~\ref{6relate}.
In 1936, Turing proposed to invent the single machine to compute any computable sequence, a concept of a "universal" computing device~\cite{Kelly2020TuringTax, Kelly2020TuringTariff}. This kind of "universal" would incur additional performance, cost, or energy overhead called "Turing Tax" -- a fundamental question that computer architects aim to reduce~\cite{Kelly2020TuringTax}.
John Gage first proposed the phrase "the network is the computer" in 1984~\cite{networkiscomputer,youtubenetwork}.
In 1985, Lewis proposed the concept of the "Internet of things" in a speech to the Congressional Black Caucus Foundation 15th Annual Legislative Weekend~\cite{iotlewis}. 
In the mid-1990s, grid computing was proposed to provide computing power, data, and software on-demand, through standardizing the protocols~\cite{foster2008cloud}. 
In 2001, EmNets~\cite{national2001embedded} were proposed and referred to as networked systems of embedded computers.
In 2005, cyber-physical systems were proposed to "bridge the cyber-world of computing and communications with the physical world"~\cite{der2005vdi,wallace2005capturing,rajkumar2010cyber}.
In 2009, Google proposed the concept of "the data center as a computer," or called warehouse-scale computers (WSCs), to efficiently deliver good levels of Internet service performance~\cite{barroso2009datacenter}.
In 2009, the Chinese Academy of Sciences predicted that human-cyber-physical ternary computing would be a development trend in the next 50 years~\cite{cas2009}.
In 2012, the "Industrial Internet of Things (IIoT)"  concept, also known as "Industrial Internet," was proposed to integrate the latest technologies, intelligence systems, and devices and apply them to the entire industrial economy~\cite{evans2012industrial,li2017industrial}.
In 2016, the director of Storage SRE at Google illustrated how they do planet-scale engineering for a planet-scale infrastructure -- keep all its services up and running and reduce the downtime~\cite{GCP2016_google}.
In 2017, Li et al. pointed out that human-cyber-physical ternary intelligence is the leading technology and the main driving force of the new economy in the next 15-20 years~\cite{guojie2017judging}.
In 2021, Mike Warren's group worked with the EC2 team, launching a virtual supercomputer in the cloud -- 4,096 EC2 instances with 172,692 cores. This run achieved 9.95 PFLOPS (actual performance), ranking at 40th on the June 2021 TOP500 list~\cite{Mike2022VSC}.
In 2021 and 2022, Wang et al. and Xu et al. pointed out "a new era of human-cyber-physical ternary computing with diverse, intelligent applications over trillions of devices"~\cite{xu2022information,wang2021building} and further proposed the concept of "Information Superbahn," to achieve high system goodput and application quality of service~\cite{xu2022information,wang2021building}.

\section{Conclusion}

%This paper first analyzes the sea change in computing, data access, and network patterns of emerging and future applications, and then points out the several orders of magnitude performance gap beyond the reach of the state-of-the-practice systems.

We call attention to the fact that more and more emerging and future applications rely heavily upon  systems consisting of Internet of Things (IoT), edges, data centers, and humans-in-the-loop.
We characterized this new class of systems and coined a new term, high fusion computers (HFCs), to describe them. Significantly different from warehouse-scale computers that non-stop serve independent concurrent user requests, HFCs directly interact with the physical world, considering humans an essential part and performing safety-critical and mission-critical operations;   their computations have intertwined dependencies between not only adjacent execution loops but also actions or decisions triggered by IoTs, edge, data centers, or humans-in-the-loops; the systems must first satisfy the accuracy metric in predicting, interpreting, or taking action before meeting the performance goal under different cases. HFCs raise severe challenges in system evaluation, design, and implementation.

We summarize several HFC challenges: organizability and manageability, collaborations between software, hardware, and people components, irreversible effect,  ecosystem wall, and effective evaluation. 
To tackle the above challenges, we propose reconstructing IoTs, edges, data centers, and humans-in-the-loop as a computer rather than a distributed system; we adopt a funclet methodology of building large systems out of smaller functions and exploring HFC design space in a structural manner. 
We will provide the first open-source implementation of the funclet architecture of HFC systems. The source code will be publicly available from the project homepage: ~\url{https://www.computercouncil.org/HFC/}.

\section*{Acknowledgement}

%We are very grateful to Dr. Ninghui Sun for his guidance.
Dr. Jianfeng Zhan contributes the concept of the IoTs, edges, data centers, and humans-in-the-loop as a computer. Dr. Yungang Bao proposes considering networking as a critical part. Dr. Jianfeng Zhan and Dr. Lei Wang co-coin the term: high fusion computers (HFCs).
Dr. Jianfeng Zhan contributes the abstract and Section 1. 
Section 2.1.1 is contributed by Dr. Jianfeng Zhan, Dr. Yunyou Huang, and Mrs. Guoxin Kang.
Section 2.1.2 is contributed by Dr. Wanling Gao, Dr. Jianfeng Zhan, and Mrs. Guoxin Kang.
Dr. Wanling Gao contributes the analysis of autonomous driving, civil aviation safety regulation, and interplanetary explorations in Table 1.
Dr. Jianfeng Zhan contributes the analysis of medical emergency management, autonomous driving, and smart defense systems in Table 1.
Dr. Yunyou Huang contributes the analysis of medical emergency management in Table 1.
Dr. Chunjie Luo and Mr. Hainan Ye contribute the analysis of Metaverse in Table 1.
Dr. Lei Wang contributes the analysis of digital twin in Table 1.
Dr. Weiping Li contributes the analysis of civil aviation safety regulation in Table 1.
%Mr. Hainan Ye contributes the analysis of 
Mrs. Guoxin Kang contributes all the analysis of data management and access patterns in Table 1.
Section 2.2.1 is contributed by Dr. Jianfeng Zhan and Dr. Wanling Gao.
Section 2.2.2 is contributed by Dr. Wanling Gao, Dr. Jianfeng Zhan, and Dr. Wenli Zhang.
Section 3 is contributed by Dr. Jianfeng Zhan and Dr. Wanling Gao. Dr. Yungang Bao contributed the organizability and manageability challenge.  
Section 4.1 is contributed by Dr. Wanling Gao and Dr. Jianfeng Zhan.
Section 4.2 is contributed by Dr. Wanling Gao, Dr. Jianfeng Zhan, and Dr. Biwei Xie.
Section 4.3 is contributed by Dr. Mingyu Chen.
Section 4.4.1 is contributed by Dr. Chen Zheng and Dr. Lei Wang.
Section 4.4.2 is contributed by Dr. Jin Xiong.
%Section 4.4.3 is contributed by Dr. Wenli Zhang.
Section 4.5.1 is contributed by Dr. Lei Wang.
Section 4.5.2 is contributed by Dr. Chunjie Luo and Dr. Biwei Xie.
Section 4.5.3 is contributed by Dr. Wanling Gao.
%Section 4.5.1 is contributed by Dr. Chunjie Luo.
%Section 4.5.2 is contributed by Dr. Wanling Gao.
%Section 4.5.3 is contributed by Fan Zhang.
%Section 4.5.4 is contributed by Dr. Lei Wang, Dr. Yisong Chang, Dr. Biwei Xie, and Hongxiao Li.
Section 5, Section 6, and Section 7 are contributed by Dr. Jianfeng Zhan and Dr. Wanling Gao.
%Section 7 is contributed by Dr. Wanling Gao.
Fig. 1, 3, 6, 7 is contributed by Dr. Wanling Gao.
Fig. 2 is contributed by Dr. Yunyou Huang.
Fig. 4 is contributed by Dr. Lei Wang, Mr. Shaopeng Dai, and Mr. Qian He.
Fig. 5 is contributed by Dr. Wanling Gao, Dr. Chunjie Luo, and Mr. Qian He.
Dr. Jianfeng Zhan, Dr. Wanling Gao, and Dr. Lei Wang proofread throughout the paper. 
We are very grateful to Dr. Sa Wang, Dr. Yisong Chang, Dr. Di Zhao, Dr. Mi Zhang, Mrs. Fan Zhang, and Mr. Hongxiao Li for their discussions and contributions.


%\bibliographystyle{plain}

%\bibliography{cas-refs}



\end{document}